# Automatic Differentiation using Operator Overloading (ADOO) for implicit resolution of hyperbolic single phase and two-phase flow models

François FRAYSSE[a] and Richard Saurel[a,b]

a: RS2N, Recherche Scientifique et Simulation Numérique, St Zacharie, France
b: LMA, Laboratoire de Mécanique et d'Acoustique, UMR 7031 AMU - CNRS - Centrale Marseille, Marseille, France

**Abstract:** Implicit time integration schemes are widely used in computational fluid dynamics numerical codes to speed-up computations. Indeed, implicit schemes usually allow for less stringent time-step stability constraints than their explicit counterpart. The derivation of an implicit scheme is however a challenging and time-consuming task, increasing substantially with the model equations complexity since this method usually requires a fairly accurate evaluation of the spatial scheme's matrix Jacobian. This article presents a flexible method to overcome the difficulties associated to the computation of the derivatives, based on the forward mode of automatic differentiation using operator overloading. Flexibility and simplicity of the method are illustrated through implicit resolution of various flow models of increasing complexity such as the compressible Euler equations, a two-phase flow model in full equilibrium (Le Martelot, et al., 2014) and a symmetric variant (Saurel , et al., 2003) of the two-phase flow model of (Baer & Nunziato, 1986) dealing with mixtures in total disequilibrium.

## I. Introduction

In the early days of numerical simulation, the industry growing needs for fast and accurate fluid flow predictions lead to numerous developments in the field of Computational Fluid Dynamics (CFD). Among them, in the aerospace community for instance, implicit time integrators arose from the motivation of decreasing the computational time required to reach steady-state, first for the Euler equations (Mulder & Van Leer, 1983) and later for the laminar compressible Navier-Stokes equations (Venkatakrishnan & Barth, 1989) and for the Reynolds-Averaged Navier-Stokes equations (Barth & Linton, 1995). Towards a better fidelity and thanks to the improvement in computational resources, the development of fast numerical schemes for unsteady flows became necessary and feasible. Taking advantage of the numerous acceleration techniques developed for the explicit time integrators solving steady-state flows, discrete unsteady equations have been solved as successive stationary problems at each time-step (Jameson, 1991). However for problems where the physical time scale is greater than the spatial scale divided by the eigenvalue, fully implicit methods are known to be more efficient (Pulliam, 1993) (Dubuc , et al., 1998).

Implicit methods for unsteady flows are known to suffer from higher diffusive and dispersive errors. Nevertheless, developments of high-order implicit methods such as those based on backward Taylor series or implicit Runge-Kutta (Gottlieb, et al., 2001) opened the way to implicit schemes applied to higher fidelity situations including direct numerical simulation (Martin & Candler, 2006) (Liu, et al., 2016).

One of the major difficulties with implicit schemes is that they require the derivatives of the spatial numerical scheme with respect to the flow variables. The complexity of the involved expressions rapidly grows with the model equations and spatial scheme sophistication such as for example, Riemann problem-based methods. This fact motivated the development of various methods requiring only a crude approximation of the Jacobian matrix such as for example the alternating direction implicit (Briley & McDonald, 1977) where the implicit operator is dimensionally split (restricted to structured grids) or the lower upper symmetric Gauss-Seidel method where the operator is decomposed in lower and upper dominant factors (Jameson & Turkel, 1981). These approximations while decreasing the development time usually severely degrade the stability of the implicit scheme yielding effective time-steps much smaller than the full unfactored form or very slow convergence of the unsteady residual.

While appealing, approximation of the matrix Jacobian through finite differentiation is an expensive strategy and subject to round-off errors not easily controlled. Krylov subspace methods for solving linear systems (Saad & Schultz, 1986) lead to matrix-free implicit algorithms known as Newton-Krylov-matrix-free. In this approach, the matrix Jacobian is not explicitly built, indeed the Krylov methods such as for example the generalized minimal residual method (GMRes) mostly only involves matrix Jacobian-vector products which can directly be approximated through finite differentiation. Nevertheless, it is well-known that the convergence rate of the GMRes method can be highly improved by the application of a preconditioner. However, forming a good preconditioner without the explicit form of the matrix is not a trivial task and the whole method is still subject to numerical error affecting both the linear and the non-linear convergence rate.

Automatic Differentiation (AD) (Wengert, 1964) (Griewank & Walther, 2000) is an algorithmic method for evaluating derivatives up to any order by propagating the chain rule to a high-level language computer program. AD combines the main advantages of symbolic and numerical differentiation without their major drawbacks. In contrast to numerical differentiation it provides exact derivatives (up to round-off error) and is less computationally demanding. AD performs on computer algorithms directly and as such is able to differentiate any complex function as well as numerical procedures such as root-finding methods. Two main approaches to AD are usually considered: AD based on source code transformation (ADSCT) and AD based on operator overloading (ADOO), the key point of the present contribution.

- ADSCT is a procedure that takes as an argument a function or a routine of a computer code, parses all basic operations and intrinsic functions, and returns another function or routine containing the tangent or adjoint derivatives which can be compiled together with the original code. This method is quite easy to apply to any programming language. The generated code complexity does not increase with respect to the original code allowing efficient compile time optimizations. This method is however rather complex to implement. Examples of codes based on this approach include ADIFOR (Bischof, et al., 1996) and TAPENADE (Hascoet & Pascual, 2013).
- ADOO determines the derivatives present in a computer code by appending a dual component to all the dependent variables. The dual variable contains the derivatives of the expression with respect to the independent variables and is propagated by applying elementary differentiation arithmetic based on the chain rule. ADOO benefits from object-oriented programming and more precisely from the ability of defining operators on user derived data-types, which is a computer language dependent feature. Programs based on this technique include ADOL-C and ADOL-F for C and FORTRAN programming language respectively (Walther & Griewank, 2012).

In this article, we will show how AD can be applied to the computation of complex matrix Jacobians and how it considerably simplifies the design of implicit schemes for sophisticated systems of equations.

The paper is organized as follows. In Section II, the derivation of a general implicit scheme targeting unsteady flows is presented. AD applied to the evaluation of the matrix Jacobian is detailed in Section III with practical examples in FORTRAN programming language. Then, various fluid flow models of increasing complexity are discretized implicitly and tested in Section IV, including the compressible Euler equations, a two-phase flow model in full equilibrium (Le Martelot, et al., 2014) and a symmetric variant (Saurel , et al., 2003) of the two-phase flow model of (Baer & Nunziato, 1986), widely used to model non-equilibrium mixtures.

## II. Implicit time integration

The fluid flow models considered in this article can be written under the following general differential form:

$$\frac{\partial \mathbf{Q}}{\partial t} + \text{div}\left(\bar{\bar{\mathbf{F}}}(\mathbf{Q})\right) + \bar{\bar{\mathbf{G}}}(\mathbf{Q}) \cdot \text{div}\left(\bar{\bar{\mathbf{H}}}(\mathbf{Q})\right) = 0 \quad (1)$$

where $\mathbf{Q}$ is the vector of conservative variables, $\text{div}\left(\bar{\bar{\mathbf{F}}}(\mathbf{Q})\right)$ the conservative part and $\bar{\bar{\mathbf{G}}}(\mathbf{Q}) \cdot \text{div}\left(\bar{\bar{\mathbf{H}}}(\mathbf{Q})\right)$ the non-conservative part of the system. The discrete finite volume form is obtained by first integrating (1) over a time-independent element $\Omega_i$ of boundary $\partial \Omega_i$ and applying the Green-Ostrogradski theorem to the conservative part:

$$\int_{\Omega_i} \frac{\partial \mathbf{Q}}{\partial t} dV_{\Omega_i} + \int_{\partial \Omega_i} \mathbf{F}(\mathbf{Q}) \cdot \mathbf{n}_{\partial \Omega_i} dS_{\partial \Omega_i} + \int_{\Omega_i} \bar{\bar{\mathbf{G}}}(\mathbf{Q}) \cdot \text{div}\left(\bar{\bar{\mathbf{H}}}(\mathbf{Q})\right) dV_{\Omega_i} = 0 \quad (2)$$

Then, assuming second order quadrature rule for the surface and volume integrals, averaging the tensor $\bar{\bar{\mathbf{G}}}(\mathbf{Q})$ over the domain $\Omega_i$ and introducing approximate Riemann fluxes $\mathbf{F}^*$ and $\mathbf{H}^*$, system (2) can be rewritten as:

$$V_i \frac{d\bar{\mathbf{Q}}_i}{dt} + \sum_{f_{ij} \in \partial \Omega_i} \mathbf{F}^*\left(g\left(\{\bar{\mathbf{Q}}_k\}_i\right), g\left(\{\bar{\mathbf{Q}}_k\}_j\right)\right) \cdot \mathbf{n}_{f_{ij}} S_{f_{ij}} + \mathbf{G}(\bar{\mathbf{Q}}_i) \sum_{f_{ij} \in \partial \Omega_i} \mathbf{H}^*\left(g\left(\{\bar{\mathbf{Q}}_k\}_i\right), g\left(\{\bar{\mathbf{Q}}_k\}_j\right)\right) \cdot \mathbf{n}_{f_{ij}} S_{f_{ij}} = 0 \quad (3)$$

In the above system, $\bar{\mathbf{Q}}_i$ is the volume average of $\mathbf{Q}$ defined as $\bar{\mathbf{Q}}_i = \frac{1}{V_i} \int_{\Omega_i} \mathbf{Q} dV_{\Omega_i}$, $f_{ij}$ is a linear surface of normal unit $\mathbf{n}_{f_{ij}}$ separating elements indexed by $i$ and $j$ and pointing towards element $j$. Function $g(\ )$ is an interpolation operator reconstructing left and right states at face $f_{ij}$ satisfying a maximum principle. This function takes as argument a set of conservative vectors forming a compact stencil centered in cell $i$ for the set $\{\bar{\mathbf{Q}}_k\}_i$ and in cell $j$ for the set $\{\bar{\mathbf{Q}}_k\}_j$.

The numerical approximation (3) with respect to non-conservative terms in the present formulation is particularly simple, but appropriate for various two-phase flow models with non-equilibrium effects (Saurel, et al., 2009) (Furfaro & Saurel, 2015). As shown in these references appropriate Riemann solvers considering these terms have to be addressed.

The implicit discrete temporal integration using Backward-Difference-Formula (BDF) under $\theta - \phi$ form gives the following non-linear system for the unknown set of vectors $\{\{\bar{\mathbf{Q}}_k\}\}_i = \bigcup_{f_{ij}} \left(\{\bar{\mathbf{Q}}_k\}_i \bigcup \{\bar{\mathbf{Q}}_k\}_i\right)$ forming the domain of dependence of cell $i$:

$$\mathbf{P}_i\left(\left\{\overline{\mathbf{Q}}_k\right\}_i^{n+1}\right) = V_i \frac{1+\phi}{\theta} \frac{\overline{\mathbf{Q}}_i^{n+1} - \overline{\mathbf{Q}}_i^n}{\Delta t} + \sum_{f_{ij} \in \partial\Omega_i} \mathbf{F}^*\left(g\left(\left\{\overline{\mathbf{Q}}_k\right\}_i^{n+1}\right), g\left(\left\{\overline{\mathbf{Q}}_k\right\}_j^{n+1}\right)\right) \cdot \mathbf{n}_{f_{ij}} S_{f_{ij}}$$

$$+ \mathbf{G}\left(\overline{\mathbf{Q}}_i^{n+1}\right) \sum_{f_{ij} \in \partial\Omega_i} \mathbf{H}^*\left(g\left(\left\{\overline{\mathbf{Q}}_k\right\}_i^{n+1}\right), g\left(\left\{\overline{\mathbf{Q}}_k\right\}_j^{n+1}\right)\right) \cdot \mathbf{n}_{f_{ij}} S_{f_{ij}}$$

$$- V_i \frac{\phi}{\theta} \frac{\overline{\mathbf{Q}}_i^n - \overline{\mathbf{Q}}_i^{n-1}}{\Delta t} = 0 \tag{4}$$

In the above system, the first order BDF method is obtained using the couple $(\theta, \phi) = (1, 0)$ and the second order BDF uses the couple $(\theta, \phi) = (1, 1/2)$.

It can be shown that BDF of order 1 (BDF1) and 2 (BDF2) are L-stable, furthermore BDF1 is unconditionally Strong Stability Preserving (SSP) for the linear case (Total Variation Diminishing-stable). However, BDF2 and all other high-order methods (>1) are only conditionally SSP, e.g. it exists a constant $c > 0 / \|\mathbf{Q}^{n+1}\| < \max\left(\|\mathbf{Q}^n\|, ..., \|\mathbf{Q}^{n+1-s}\|\right)$, for $\Delta t < c\Delta t_{EE}$ where $\Delta t_{EE}$ is the maximum time step for TVD Explicit Euler integration. This constant c is not easily bounded in the general case, and often $\Delta t$ is bounded by criteria based on relevant physics of the problem under study.

System (4) written for all cells of the mesh represents a coupled non-linear set of equations for the next time-step unknown vector $\overline{\mathbf{Q}}^{n+1}$. Its solution is approximated by the iterative Newton-Raphson (NR) method. An NR method iteration is written as $\overline{\mathbf{Q}}^{r+1} = \overline{\mathbf{Q}}^r + \Delta\overline{\mathbf{Q}}^r$ where the increment $\Delta\overline{\mathbf{Q}}^r$ is given by the solution of the linear system $\frac{\partial \mathbf{P}}{\partial \overline{\mathbf{Q}}^r} \Delta\overline{\mathbf{Q}}^r = -\mathbf{P}\left(\overline{\mathbf{Q}}^r\right)$. At convergence of the NR method, e.g. when $\left\|\frac{\Delta\overline{\mathbf{Q}}^r}{\overline{\mathbf{Q}}^n}\right\| < \varepsilon_{NL}$, where $\varepsilon_{NL}$ is a tolerance parameter, the solution at the next time-step is obtained: $\overline{\mathbf{Q}}^{n+1} = \lim_{\left\|\frac{\Delta\overline{\mathbf{Q}}^r}{\overline{\mathbf{Q}}^n}\right\| \to \varepsilon_{NL}} \overline{\mathbf{Q}}^{r+1}$.

The NR method thus requires a sequence of linear problems to solve each time-step involving the derivative of the function $\mathbf{P}$ with respect to the conservative variables $\frac{\partial \mathbf{P}}{\partial \overline{\mathbf{Q}}^r}$. In compact form, the Newton iteration is obtained through the solution of the following linear system:

$$\left(\frac{1+\phi}{\theta} \frac{V_i}{\Delta t} \mathbf{I}_d + \mathbf{J}\right) \Delta\overline{\mathbf{Q}}_i^r = -\sum_{f_{ij} \in \partial\Omega_i} \mathbf{F}^*\left(g\left(\left\{\overline{\mathbf{Q}}_k\right\}_i^r\right), g\left(\left\{\overline{\mathbf{Q}}_k\right\}_j^r\right)\right) \cdot \mathbf{n}_{f_{ij}} S_{f_{ij}}$$

$$- \mathbf{G}\left(\overline{\mathbf{Q}}_i^r\right) \sum_{f_{ij} \in \partial\Omega_i} \mathbf{H}^*\left(g\left(\left\{\overline{\mathbf{Q}}_k\right\}_i^r\right), g\left(\left\{\overline{\mathbf{Q}}_k\right\}_j^r\right)\right) \cdot \mathbf{n}_{f_{ij}} S_{f_{ij}}$$

$$- V_i \frac{1+\phi}{\theta} \frac{\overline{\mathbf{Q}}_i^s - \overline{\mathbf{Q}}_i^n}{\Delta t} + V_i \frac{\phi}{\theta} \frac{\overline{\mathbf{Q}}_i^n - \overline{\mathbf{Q}}_i^{n-1}}{\Delta t} \tag{5}$$

where the matrix Jacobian J gathers all the conservative and non-conservative flux derivatives present in the implicit operator $\frac{\partial \mathbf{P}}{\partial \overline{\mathbf{Q}}^r}$. In case of unstructured grids, J is a sparse non-symmetric block-matrix whose level of fill-in depends on the union of all spatial domains of dependence which is equal to $\mathrm{card}\left(\bigcup_i \{\{\overline{\mathbf{Q}}_k\}\}_i\right)$. Let us consider $\overline{\mathbf{Q}}_m \subset \{\{\overline{\mathbf{Q}}_k\}\}_i$, the block matrix $\mathbf{J}_{im}$ of $\mathbf{J}$ is related to $\dfrac{\partial \mathbf{P}}{\partial \overline{\mathbf{Q}}^r}$ through the following equalities:

$$\text{if } m \neq i, \left.\frac{\partial \mathbf{P}_i}{\partial \overline{\mathbf{Q}}_m^r}\right|_{\{\overline{\mathbf{Q}}_k\}_i^r} = \sum_{f_{ij} \in \partial \Omega_i} \left.\frac{\partial \mathbf{F}^*}{\partial \overline{\mathbf{Q}}_m^r}\right|_{\left(g\left(\{\overline{\mathbf{Q}}_k\}_i^r\right), g\left(\{\overline{\mathbf{Q}}_k\}_j^r\right)\right)} \cdot \mathbf{n}_{f_{ij}} S_{f_{ij}} + \mathbf{G}\left(\overline{\mathbf{Q}}_i^r\right) \sum_{f_{ij} \in \partial \Omega_i} \left.\frac{\partial \mathbf{H}^*}{\partial \overline{\mathbf{Q}}_m^r}\right|_{\left(g\left(\{\overline{\mathbf{Q}}_k\}_i^r\right), g\left(\{\overline{\mathbf{Q}}_k\}_j^r\right)\right)} \cdot \mathbf{n}_{f_{ij}} S_{f_{ij}}$$

$$= \mathbf{J}_{im}$$

$$\text{if } m = i, \left.\frac{\partial \mathbf{P}_i}{\partial \overline{\mathbf{Q}}_m^r}\right|_{\{\overline{\mathbf{Q}}_k\}_i^r} = \sum_{f_{ij} \in \partial \Omega_i} \left.\frac{\partial \mathbf{F}^*}{\partial \overline{\mathbf{Q}}_m^r}\right|_{\left(g\left(\{\overline{\mathbf{Q}}_k\}_i^r\right), g\left(\{\overline{\mathbf{Q}}_k\}_j^r\right)\right)} \cdot \mathbf{n}_{f_{ij}} S_{f_{ij}} + \mathbf{G}\left(\overline{\mathbf{Q}}_i^r\right) \sum_{f_{ij} \in \partial \Omega_i} \left.\frac{\partial \mathbf{H}^*}{\partial \overline{\mathbf{Q}}_m^r}\right|_{\left(g\left(\{\overline{\mathbf{Q}}_k\}_i^r\right), g\left(\{\overline{\mathbf{Q}}_k\}_j^r\right)\right)} \cdot \mathbf{n}_{f_{ij}} S_{f_{ij}}$$

$$+ \left.\frac{\partial \mathbf{G}}{\partial \overline{\mathbf{Q}}_m^r}\right|_{\overline{\mathbf{Q}}_i^r} \sum_{f_{ij} \in \partial \Omega_i} \mathbf{H}^*\left(g\left(\{\overline{\mathbf{Q}}_k\}_i^r\right), g\left(\{\overline{\mathbf{Q}}_k\}_j^r\right)\right) \cdot \mathbf{n}_{f_{ij}} S_{f_{ij}} + V_i \frac{1+\phi}{\theta} \mathbf{I}_d$$

$$= \mathbf{J}_{im} + V_i \frac{1+\phi}{\theta} \mathbf{I}_d$$

Performing a single Newton iteration results in the so-called linearized implicit method, which requires only one linear system resolution per time-step. Such approximation allows a large gain in computational speed, but its accuracy is clearly restricted to the case where the magnitude of the unsteady residual is lower than the temporal scheme's truncation error: $\left\|\dfrac{\Delta \overline{\mathbf{Q}}^{r=1}}{\overline{\mathbf{Q}}^n}\right\| \approx O\left(\Delta t^k\right)$. Nevertheless, this condition is fulfilled under the following situations:

- Weak temporal non-linearity;
- Accurate estimation of the initial guess in the Newton's algorithm;
- Accurate evaluation of the Jacobian matrix.

Backward difference formula of order two is not a self-starting method as it requires the solution at time n-1. Consequently, the numerical integration for the first time-step needs another discretization scheme. In this paper, the first-order backward difference formula is used to start the numerical integration.

The second implicit scheme considered in this paper is a semi-implicit scheme of the Runge-Kutta family. More precisely, the two-stage second-order Strong Stability Preserving Singly Diagonally Implicit Runge Kutta scheme (SSPSDIRK-2) (Kennedy & Carpenter, 2016). This is a multi-step method and as such is self-starting. The semi-implicit terminology refers to the fact that the Runge-Kutta stages are decoupled from each other, resulting in the present case in two sequential non-linear equations per time-step. This scheme can be written in the following compact form:

$$\left(\frac{1+\phi}{\theta}\frac{V_i}{\Delta t}\mathbf{I}_d + a_{11}\mathbf{J}\right)\Delta\overline{\mathbf{Q}}_i^{r,1} = -a_{11}\left(\sum_{f_{ij}\in\partial\Omega_i}\mathbf{F}^*\left(g\left(\{\overline{\mathbf{Q}}_k\}_i^{r,1}\right),g\left(\{\overline{\mathbf{Q}}_k\}_j^{r,1}\right)\right)\cdot\mathbf{n}_{f_{ij}}S_{f_{ij}} + \mathbf{G}\left(\overline{\mathbf{Q}}_i^{r,1}\right)\sum_{f_{ij}\in\partial\Omega_i}\mathbf{H}^*\left(g\left(\{\overline{\mathbf{Q}}_k\}_i^{r,1}\right),g\left(\{\overline{\mathbf{Q}}_k\}_j^{r,1}\right)\right)\cdot\mathbf{n}_{f_{ij}}S_{f_{ij}}\right)$$

$$-V_i\frac{\overline{\mathbf{Q}}_i^{r,1} - \overline{\mathbf{Q}}_i^n}{\Delta t}$$

$$\text{with}\quad \overline{\mathbf{Q}}_i^{1,1} = \overline{\mathbf{Q}}_i^n, \overline{\mathbf{Q}}_i^{r+1,1} = \overline{\mathbf{Q}}_i^{r,1} + \Delta\overline{\mathbf{Q}}_i^{r,1}, \overline{\mathbf{Q}}_i^1 = \lim_{\Delta\overline{\mathbf{Q}}_i^{r,1}\to 0}\overline{\mathbf{Q}}_i^{r+1,1}$$

<u>Stage 1</u>

$$\left(\frac{1+\phi}{\theta}\frac{V_i}{\Delta t}\mathbf{I}_d + a_{22}\mathbf{J}\right)\Delta\overline{\mathbf{Q}}_i^{r,2} = -a_{11}\left(\sum_{f_{ij}\in\partial\Omega_i}\mathbf{F}^*\left(g\left(\{\overline{\mathbf{Q}}_k\}_i^{r,2}\right),g\left(\{\overline{\mathbf{Q}}_k\}_j^{r,2}\right)\right)\cdot\mathbf{n}_{f_{ij}}S_{f_{ij}} + \mathbf{G}\left(\overline{\mathbf{Q}}_i^{r,2}\right)\sum_{f_{ij}\in\partial\Omega_i}\mathbf{H}^*\left(g\left(\{\overline{\mathbf{Q}}_k\}_i^{r,2}\right),g\left(\{\overline{\mathbf{Q}}_k\}_j^{r,2}\right)\right)\cdot\mathbf{n}_{f_{ij}}S_{f_{ij}}\right)$$

$$-a_{21}\left(\sum_{f_{ij}\in\partial\Omega_i}\mathbf{F}^*\left(g\left(\{\overline{\mathbf{Q}}_k\}_i^1\right),g\left(\{\overline{\mathbf{Q}}_k\}_j^1\right)\right)\cdot\mathbf{n}_{f_{ij}}S_{f_{ij}} + \mathbf{G}\left(\overline{\mathbf{Q}}_i^1\right)\sum_{f_{ij}\in\partial\Omega_i}\mathbf{H}^*\left(g\left(\{\overline{\mathbf{Q}}_k\}_i^1\right),g\left(\{\overline{\mathbf{Q}}_k\}_j^1\right)\right)\cdot\mathbf{n}_{f_{ij}}S_{f_{ij}}\right)$$

$$-V_i\frac{\overline{\mathbf{Q}}_i^{r,2} - \overline{\mathbf{Q}}_i^n}{\Delta t}$$

$$\text{with}\quad \overline{\mathbf{Q}}_i^{1,2} = \overline{\mathbf{Q}}_i^1, \overline{\mathbf{Q}}_i^{r+1,2} = \overline{\mathbf{Q}}_i^{r,2} + \Delta\overline{\mathbf{Q}}_i^{r,2}, \overline{\mathbf{Q}}_i^2 = \lim_{\Delta\overline{\mathbf{Q}}_i^{r,2}\to 0}\overline{\mathbf{Q}}_i^{r+1,2}$$

<u>Stage 2</u>

$$\overline{\mathbf{Q}}_i^{n+1} = \overline{\mathbf{Q}}_i^n + \frac{\Delta t}{V_i}\sum_{s=1}^{2}b_s\left(\sum_{f_{ij}\in\partial\Omega_i}\mathbf{F}^*\left(g\left(\{\overline{\mathbf{Q}}_k\}_i^s\right),g\left(\{\overline{\mathbf{Q}}_k\}_j^s\right)\right)\cdot\mathbf{n}_{f_{ij}}S_{f_{ij}} + \mathbf{G}\left(\overline{\mathbf{Q}}_i^s\right)\sum_{f_{ij}\in\partial\Omega_i}\mathbf{H}^*\left(g\left(\{\overline{\mathbf{Q}}_k\}_i^s\right),g\left(\{\overline{\mathbf{Q}}_k\}_j^s\right)\right)\cdot\mathbf{n}_{f_{ij}}S_{f_{ij}}\right)$$

with

$$a = \begin{pmatrix} 1-\frac{\sqrt{2}}{2} & 0 \\ \sqrt{2}-1 & 1-\frac{\sqrt{2}}{2} \end{pmatrix}, b = \begin{pmatrix} \frac{1}{2} \\ \frac{1}{2} \end{pmatrix}$$

For moderately and highly unsteady flows, forming an accurate Jacobian matrix is of primary importance to reach good convergence rate in the Newton's method (up to quadratic).

## III. Evaluation of the Jacobian matrix

### 1. Introduction

Two main strategies are usually followed to evaluate the Jacobian matrix of the spatial numerical scheme:

- Analytic differentiation. This approach involves both hand and symbolic differentiation of the numerical flux and overall scheme. The main advantage of this approach is the full determination and control of the elements of the matrix resulting generally in an accurate and efficient Jacobian evaluation. Various drawbacks are however present. First, the complexity of the differentials grows rapidly with the model and numerical scheme sophistication making this method very error-prone. The level of flexibility is very low since any slight modification in the model or in the numerical scheme needs an adaptation of the Jacobian matrix. Further difficulties arise when the numerical scheme involves root-finding algorithms such as sophisticated boundary conditions and non-explicit equations of state (EOS). Symbolic differentiation is somehow more flexible but often at the cost of rather complicated expressions thus resulting in a larger computational overhead.

- Numerical differentiation. This method is based on finite differencing, and as such is not an exact method. The aim is to approximate the implicit operator as

$$\frac{\partial \mathbf{P}}{\partial \overline{\mathbf{Q}}^r} \approx \frac{\mathbf{P}(\overline{\mathbf{Q}}^r + \varepsilon_{FD}) - \mathbf{P}(\overline{\mathbf{Q}}^r)}{\varepsilon_{FD}}$$

with $\varepsilon_{FD}$ a numerical tolerance. While offering a maximum flexibility and ease of implementation, this approach has two major drawbacks. First, the computational cost is relatively high as one evaluation of the function P is necessary for each degree of freedom. The second drawback lies in the calibration of $\varepsilon_{FD}$. Indeed, its value is of fundamental importance as too high or too small values yield inaccurate derivative approximations because of truncation and round-off errors respectively (Dennis & Schnabel, 1983). This approach gained however a lot of interest and success when used together with Krylov subspace methods as a solver for the linear problems (Brown & Saad, 1990). Indeed, in these linear solvers the implicit operator only appears as a product with the solution increment $\Delta \overline{\mathbf{Q}}^r$. Consequently, the full matrix does not need to be evaluated, instead the matrix-vector product is approximated through finite differencing:

$$\frac{\partial \mathbf{P}}{\partial \mathbf{Q}^r} \Delta \mathbf{Q}^r \approx \frac{\mathbf{P}(\mathbf{Q}^r + \varepsilon_{FD} \Delta \mathbf{Q}^r) - \mathbf{P}(\mathbf{Q}^r)}{\varepsilon_{FD}}$$

While being attractive in terms of computational speed with respect to the full matrix approximation, this method still suffers of non-trivial estimate of $\varepsilon_{FD}$. An inaccurate approximation of the implicit operator can severely degrade the convergence rate of the Newton's method or the accuracy of the linearized implicit method. Furthermore, it is well known that the convergence rate of the iterative Krylov subspace linear solvers degrades rapidly when the diagonal dominance decreases (e.g. when the time-step increases). Convergence rate of the linear solver is usually significantly increased by the application of a preconditioner which tends to decrease the spectral radius of the implicit operator. Efficient fully matrix-free preconditioning is nonetheless not trivial and is still an active research area.

2. **Automatic differentiation**

As we shall see in the following, ADOO can be implemented in a straightforward manner into an existing FORTRAN code. Let us first consider the following simple example:

$$f(x) = \sin(x) + x^2$$
$$f'(x) = \cos(x) + 2x$$

The first step consists in replacing the real variables $x$ and $f$ by two objects $x_{AD}$ and $f_{AD}$ having two components $\begin{bmatrix} v \\ dv \end{bmatrix}$: $v$ being the value of the variable itself and $dv$ the value of its derivative. The FORTRAN 90 standard allows the construction of such objects, more commonly called derived data types (DDT). The FORTRAN 2003 standard allows each component of a DDT, and in particular $dv$, to be an array which gives a greater flexibility in handling derivatives with respect to various independent variables. The evaluation of the derivative following the ADOO approach is performed in three steps:

$$x_{AD} = \underbrace{\begin{bmatrix} x \\ 1 \end{bmatrix}}_{\text{Initialization}} \Rightarrow f_{AD}(x_{AD}) = \underbrace{\sin_{AD}\left(\begin{bmatrix} x \\ 1 \end{bmatrix}\right) +_{AD} \begin{bmatrix} x \\ 1 \end{bmatrix}^{2_{AD}}}_{\text{Operator overloading}} = \underbrace{\begin{bmatrix} \sin(x) \\ \cos(x) \end{bmatrix} +_{AD} \begin{bmatrix} x^2 \\ 2x \end{bmatrix} = \begin{bmatrix} \sin(x) + x^2 \\ \cos(x) + 2x \end{bmatrix} \equiv \begin{bmatrix} f(x) \\ f'(x) \end{bmatrix}}_{\text{Propagation}}$$

- Initialization. It consists in copying the value of $x$ in the component $v$ of DDT $x_{AD}$ and initializing the component $dv$ to 1, indeed in this case $dv = \dfrac{dx}{dx} = 1$
- Operator overloading. A DDT is a construct defined by the programmer. As such, no rules are predefined for operations involving one or more DDTs and they need to be coded. Fortunately, the number of basic operations and intrinsic functions is finite, and their coding can be done once and for all. In the above example, rules must be defined for the operators $\sin_{AD}, +_{AD}, (\ )^{2_{AD}}$.
- Propagation. Once all needed operators and intrinsic functions are overloaded they can be applied to any DDT and the derivatives then propagate through the chain rule.

To better understand how ADOO works, let us consider the one-dimensional compressible Euler equations closed by the ideal gas equation of state:

$$\frac{\partial \mathbf{Q}}{\partial t} + \frac{\partial \mathbf{F}(\mathbf{Q})}{\partial x} = 0 \text{ with } \mathbf{Q} = \begin{pmatrix} \rho \\ \rho u \\ \rho E \end{pmatrix}, \mathbf{F}(\mathbf{Q}) = \begin{pmatrix} \rho u \\ \rho u^2 + p \\ u(\rho E + p) \end{pmatrix} \text{ and } E = e(\rho, p) + \frac{1}{2} u^2 = \frac{p}{\rho(\gamma - 1)} + \frac{1}{2} u^2$$

where $\rho, u, p, E, e, \gamma$ denote the density, the velocity, the pressure, the total energy, the internal energy and the polytropic gas constant respectively. Let us assume a BDF1 numerical scheme in time and a first order finite volume discretization in space on a uniform grid indexed by $i$ and of spacing $\Delta x$. A single Newton iteration (linearized implicit) simplifies the non-linear system (4) to the following tridiagonal linear system to be solved at each time-step:

$$\left( \frac{\Delta x}{\Delta t} \mathbf{I}_d + \frac{\partial \mathbf{F}_{i,i+1}^{*n}}{\partial \overline{\mathbf{Q}}_i^n} - \frac{\partial \mathbf{F}_{i-1,i}^{*n}}{\partial \overline{\mathbf{Q}}_i^n} \right) \Delta \overline{\mathbf{Q}}_i^n - \frac{\partial \mathbf{F}_{i-1,i}^{*n}}{\partial \overline{\mathbf{Q}}_{i-1}^n} \Delta \overline{\mathbf{Q}}_{i-1}^n + \frac{\partial \mathbf{F}_{i,i+1}^{*n}}{\partial \overline{\mathbf{Q}}_{i+1}^n} \Delta \overline{\mathbf{Q}}_{i+1}^n = -\left[ \mathbf{F}_{i,i+1}^{*n} - \mathbf{F}_{i-1,i}^{*n} \right]$$

$$\text{with } \mathbf{F}_{i,i+1}^{*n} \equiv \mathbf{F}^*\left( \overline{\mathbf{Q}}_i^n, \overline{\mathbf{Q}}_{i+1}^n \right) \text{ and } \Delta \overline{\mathbf{Q}}_i^n = \overline{\mathbf{Q}}_i^{n+1} - \overline{\mathbf{Q}}_i^n \qquad (6)$$

Let us approximate the intercell fluxes by the two-waves Riemann solver of Rusanov (Rusanov, 1962) which has a relatively low algorithmic complexity:

$$\mathbf{F}^*\left( \overline{\mathbf{Q}}_i^n, \overline{\mathbf{Q}}_{i+1}^n \right) = \frac{1}{2} \left( \mathbf{F}\left( \overline{\mathbf{Q}}_i^n \right) + \mathbf{F}\left( \overline{\mathbf{Q}}_{i+1}^n \right) - S_{MAX} \left( \overline{\mathbf{Q}}_{i+1}^n - \overline{\mathbf{Q}}_i^n \right) \right)$$

$$\text{with } S_{MAX} = MAX\left( |u_i^n| + c_i^n, |u_{i+1}^n| + c_{i+1}^n \right) \text{ and } c_i^n = \sqrt{\frac{\gamma p_i^n}{\rho_i^n}}$$

Let us focus on the evaluation of the local Jacobian matrix $\dfrac{\partial \mathbf{F}^{*n}_{i,i+1}}{\partial \overline{\mathbf{Q}}^{n}_{i}}$ appearing in (6) in the special case $S_{max} = u^n_i + c^n_i, u^n_i > 0$. Omitting the time superscript $n$ and average symbol $\overline{\phantom{x}}$, the exact Jacobian obtained by hand differentiation is given by:

$$\dfrac{\partial \mathbf{F}^{*n}_{i,i+1}}{\partial \overline{\mathbf{Q}}^{n}_{i}} = \begin{pmatrix} \dfrac{u_i+c_i}{2} - \left(\dfrac{\gamma(\gamma-1)}{4\rho_i c_i}\left(u_i^2-E_i\right)-\dfrac{u_i}{2\rho_i}\right)(\rho_{i+1}-\rho_i) & \dfrac{1}{2}-\left(\dfrac{1}{2\rho_i}-\dfrac{\gamma(\gamma-1)}{4\rho_i c_i}u_i\right)(\rho_{i+1}-\rho_i) & -\dfrac{\gamma(\gamma-1)}{4\rho_i c_i}(\rho_{i+1}-\rho_i) \\ \dfrac{\gamma-3}{4}u_i^2 - \left(\dfrac{\gamma(\gamma-1)}{4\rho_i c_i}\left(u_i^2-E_i\right)-\dfrac{u_i}{2\rho_i}\right)(\rho_{i+1}u_{i+1}-\rho_i u_i) & \dfrac{3-\gamma}{2}u_i+u_i+c_i-\left(\dfrac{1}{2\rho_i}-\dfrac{\gamma(\gamma-1)}{4\rho_i c_i}u_i\right)(\rho_{i+1}u_{i+1}-\rho_i u_i) & \dfrac{\gamma-1}{2}-\dfrac{\gamma(\gamma-1)}{4\rho_i c_i}(\rho_{i+1}u_{i+1}-\rho_i u_i) \\ \dfrac{(\gamma-1)u_i^3-\gamma u_i E_i}{2}-\left(\dfrac{\gamma(\gamma-1)}{4\rho_i c_i}\left(u_i^2-E_i\right)-\dfrac{u_i}{2\rho_i}\right)(\rho_{i+1}E_{i+1}-\rho_i E_i) & \dfrac{2\gamma E_i - 3(\gamma-1)u_i^2}{4}-\left(\dfrac{1}{2\rho_i}-\dfrac{\gamma(\gamma-1)}{4\rho_i c_i}u_i\right)(\rho_{i+1}E_{i+1}-\rho_i E_i) & \dfrac{(\gamma+1)u_i+c_i}{2}-\dfrac{\gamma(\gamma-1)}{4\rho_i c_i}(\rho_{i+1}E_{i+1}-\rho_i E_i) \end{pmatrix} \quad (7)$$

In order to compare this hand differentiated Jacobian to the one obtained by ADOO in FORTRAN language, the first step consists in defining a DDT as shown in Figure 1.

```
PUBLIC :: AutoDiffType
TYPE AutoDiffType
  REAL(KIND=RP)                :: v
  REAL(KIND=RP)                :: dv(NCons,Nderiv)
END TYPE AutoDiffType
```

**Figure 1: FORTRAN definition of a derived data type.**

Note the dimension of the component $dv$ which corresponds to the total number of independent variables. In the case of one-dimensional Euler equations NCons represents the number of conservative variables which is equal to 3 and Nderiv the number of conservative vectors the intercell flux depends on. Nderiv is equal to 2 in the case of first order spatial discretization $\left(\overline{\mathbf{Q}}^{n}_{i}, \overline{\mathbf{Q}}^{n}_{i+1}\right)$.

The second step consists in overloading operators and intrinsic functions needed to compute the intercell flux in order to handle computations involving DDTs. Examples are given in Figure 2.

```
PURE ELEMENTAL SUBROUTINE d_assign_d(d1,d2)
TYPE(AutoDiffType), INTENT(OUT) :: d1
TYPE(AutoDiffType), INTENT( IN) :: d2

d1%v  = d2%v
d1%dv = d2%dv

END SUBROUTINE d_assign_d

PURE ELEMENTAL SUBROUTINE d_assign_r(d,r)
TYPE(AutoDiffType), INTENT(OUT) :: d
REAL(KIND=RP)     , INTENT( IN) :: r

d%v  = r
d%dv = ZERO

END SUBROUTINE d_assign_r

PURE ELEMENTAL SUBROUTINE r_assign_d(r,d)
REAL(KIND=RP)     , INTENT(OUT) :: r
TYPE(AutoDiffType), INTENT( IN) :: d

r = d%v

END SUBROUTINE r_assign_d
```

```
PURE ELEMENTAL FUNCTION d_times_d(d1, d2)
TYPE(AutoDiffType), INTENT(IN) :: d1, d2
TYPE(AutoDiffType)             :: d_times_d

d_times_d%v  = d1%v*d2%v
d_times_d%dv = d1%v*d2%dv + d2%v*d1%dv

END FUNCTION d_times_d

PURE ELEMENTAL FUNCTION d_times_r(d, r)
TYPE(AutoDiffType), INTENT(IN) :: d
REAL(KIND=RP)     , INTENT(IN) :: r
TYPE(AutoDiffType)             :: d_times_r

d_times_r%v  = r*d%v
d_times_r%dv = r*d%dv

END FUNCTION d_times_r

PURE ELEMENTAL FUNCTION r_times_d(r, d)
TYPE(AutoDiffType), INTENT(IN) :: d
REAL(KIND=RP)     , INTENT(IN) :: r
TYPE(AutoDiffType)             :: r_times_d

r_times_d%v  = r*d%v
r_times_d%dv = r*d%dv

END FUNCTION r_times_d
```

```
PURE ELEMENTAL FUNCTION SIN_d(d)
TYPE(AutoDiffType), INTENT(IN) :: d
TYPE(AutoDiffType)             :: SIN_d

SIN_d%v  = SIN(d%v)
SIN_d%dv = d%dv * COS(d%v)

END FUNCTION SIN_d

PURE ELEMENTAL FUNCTION COS_d(d)
TYPE(AutoDiffType), INTENT(IN) :: d
TYPE(AutoDiffType)             :: COS_d

COS_d%v  = COS(d%v)
COS_d%dv = -d%dv * SIN(d%v)

END FUNCTION COS_d

PURE ELEMENTAL FUNCTION EXP_d(d)
TYPE(AutoDiffType), INTENT(IN) :: d
TYPE(AutoDiffType)             :: EXP_d

EXP_d%v  = EXP(d%v)
EXP_d%dv = d%dv*EXP(d%v)

END FUNCTION EXP_d
```

**Figure 2: Examples of operator and function overloading. These functions allow the use of basic arithmetic with REAL type, AutoDiffType or a mix of both.**

Let us focus on the function d_times_d described in Figure 2. This function takes as arguments two DDTs and returns a DDT whose component $v$ holds the product of the values and the component $dv$ holds the conventional derivative of the product. Note the keyword ELEMENTAL, which is a FORTRAN 95 standard and allows to call this function with d1 and d2 being scalar DDTs or arrays of DDTs in a transparent way. The functions d_times_r and r_times_d are necessary when multiplying a DDT by a REAL constant. Note that similar rules may be defined for the product of an INTEGER by a DDT and vice versa.

At this point, all basic arithmetic operators and functions in the code involving DDTs would have to be rewritten to use the aforementioned routines. Fortunately, in FORTRAN, an automatic selection of the right function can be achieved through interfacing, see example in Figure 3.

```
PUBLIC :: ASSIGNMENT(=)
INTERFACE ASSIGNMENT(=)
  MODULE PROCEDURE d_assign_d
  MODULE PROCEDURE d_assign_r
  MODULE PROCEDURE r_assign_d
END INTERFACE

PUBLIC :: OPERATOR(*)
INTERFACE OPERATOR(*)
  MODULE PROCEDURE d_times_d
  MODULE PROCEDURE d_times_r
  MODULE PROCEDURE r_times_d
END INTERFACE
```

```
PUBLIC :: SIN
INTERFACE SIN
  MODULE PROCEDURE SIN_d
END INTERFACE

PUBLIC :: COS
INTERFACE COS
  MODULE PROCEDURE COS_d
END INTERFACE

PUBLIC :: EXP
INTERFACE EXP
  MODULE PROCEDURE EXP_d
END INTERFACE
```

**Figure 3: Example of operator and function overloading. Interface blocks allowing the use of the symbols =,* and the intrinsic functions SIN,COS and EXP whether the arguments are of type REAL or AutoDiffType or a mix of both.**

The selection is based on the type of the argument at compile time.

Once all required operators and functions are properly overloaded, the last step is to initialize the component $dv$ of the independent variables to the identity matrix $\left( \frac{\partial \overline{\mathbf{Q}}_i^n}{\partial \overline{\mathbf{Q}}_i^n} = \frac{\partial \overline{\mathbf{Q}}_{i+1}^n}{\partial \overline{\mathbf{Q}}_{i+1}^n} = \mathbf{I}_d \right)$ as shown in Figure 4.

```
SUBROUTINE InitDerivatives(VcL,VcR)
TYPE(AutoDiffType),INTENT(INOUT) :: VcL(:),VcR(:)

VcL(VC_Rho)%dv(:,:)=ZERO
VcL(VC_RhoU)%dv(:,:)=ZERO
VcL(VC_RhoE)%dv(:,:)=ZERO

VcR(VC_Rho)%dv(:,:)=ZERO
VcR(VC_RhoU)%dv(:,:)=ZERO
VcR(VC_RhoE)%dv(:,:)=ZERO

VcL(VC_Rho)%dv(VC_Rho,LEFT)=ONE
VcL(VC_RhoU)%dv(VC_RhoU,LEFT)=ONE
VcL(VC_RhoE)%dv(VC_RhoE,LEFT)=ONE

VcR(VC_Rho)%dv(VC_Rho,RIGHT)=ONE
VcR(VC_RhoU)%dv(VC_RhoU,RIGHT)=ONE
VcR(VC_RhoE)%dv(VC_RhoE,RIGHT)=ONE

END SUBROUTINE InitDerivatives
```

**Figure 4: Initialization of the derivatives of the left and right vectors of conservative variables prior application of the chain rule.**

Finally, a call to the routine computing the Rusanov flux (Figure 5) will automatically compute all the derivatives of the flux with respect to the left and to the right vectors of conservative variables: the v component of the DDT variable "Flux(:)" holds all the necessary derivatives. For example, $\frac{\partial F^*_{\rho_{L,R}}}{\partial \rho_L}$ is stored in Flux(VC_Rho)%dv(VC_Rho,LEFT) and $\frac{\partial F^*_{\rho_{L,R}}}{\partial \rho_R}$ is stored in Flux(VC_Rho)%dv(VC_Rho,RIGHT).

```
SUBROUTINE RusanovFlux(VcL,VcR,Flux)
TYPE(AutoDiffType),INTENT(IN)::VcL(:),VcR(:)
TYPE(AutoDiffType),INTENT(OUT)::Flux(:)
TYPE(AutoDiffType)::rhoL,uL,pL,EL,cL
TYPE(AutoDiffType)::rhoR,uR,pR,ER,cR
TYPE(AutoDiffType)::Smax
TYPE(AutoDiffType),DIMENSION(Ncons)::fluxL,fluxR

rhoL=VcL(VC_Rho)
uL=VcL(VC_RhoU)/rhoL
EL=VcL(VC_RhoE)/rhoL
pL=rhoL*(GAMMA-ONE)*(EL-HALF*uL**2)

rhoR=VcR(VC_Rho)
uR=VcR(VC_RhoU)/rhoR
ER=VcR(VC_RhoE)/rhoR
pR=rhoR*(GAMMA-ONE)*(ER-HALF*uR**2)

cL=SQRT(GAMMA*pL/rhoL)
cR=SQRT(GAMMA*pR/rhoR)

Smax=MAX(ABS(uL)+cL,ABS(uR)+cR)

fluxL=[rhoL*uL,rhoL*uL**2+pL,uL*(rhoL*EL+pL)]
fluxR=[rhoR*uR,rhoR*uR**2+pR,uR*(rhoR*ER+pR)]

Flux(:)=HALF*(fluxL(:)+fluxR(:)-Smax*(VcR(:)-VcL(:)))

END SUBROUTINE RusanovFlux
```

**Figure 5: FORTRAN routine for the computation of the Rusanov flux as well as all its derivatives with respect to the left and right vectors of conservative variables. It is worth to note that except the declaration of the variables where TYPE(AutoDiffType) is inserted, the rest of the subroutine corresponds to the conventional one used in the explicit version of the method.**

Without loss of generality, let us detail the computation of $\dfrac{\partial F^*_{\rho_{L,R}}}{\partial \rho_L}$. After the initialization step detailed in Figure 4, the left and right $dv$ components of the conservative vectors $V_{c_L}, V_{c_R}$ are filled with the following values:

$$V_{c_L}(VC\_Rho) \Rightarrow \begin{cases} v:\rho_L \\ dv: \begin{matrix}L\\R\end{matrix}\begin{bmatrix}1 & 0 & 0 \\ 0 & 0 & 0\end{bmatrix}^T \end{cases} \quad V_{c_R}(VC\_Rho) \Rightarrow \begin{cases} v:\rho_R \\ dv: \begin{matrix}L\\R\end{matrix}\begin{bmatrix}0 & 0 & 0 \\ 1 & 0 & 0\end{bmatrix}^T \end{cases}$$

$$V_{c_L}(VC\_RhoU) \Rightarrow \begin{cases} v:\rho_L u_L \\ dv: \begin{matrix}L\\R\end{matrix}\begin{bmatrix}0 & 1 & 0 \\ 0 & 0 & 0\end{bmatrix}^T \end{cases} \quad V_{c_R}(VC\_RhoU) \Rightarrow \begin{cases} v:\rho_R u_R \\ dv: \begin{matrix}L\\R\end{matrix}\begin{bmatrix}0 & 0 & 0 \\ 0 & 1 & 0\end{bmatrix}^T \end{cases}$$

$$V_{c_L}(VC\_RhoE) \Rightarrow \begin{cases} v:\rho_L E_L \\ dv: \begin{matrix}L\\R\end{matrix}\begin{bmatrix}0 & 0 & 1 \\ 0 & 0 & 0\end{bmatrix}^T \end{cases} \quad V_{c_R}(VC\_RhoE) \Rightarrow \begin{cases} v:\rho_R E_R \\ dv: \begin{matrix}L\\R\end{matrix}\begin{bmatrix}0 & 0 & 0 \\ 0 & 0 & 1\end{bmatrix}^T \end{cases}$$

In the Rusanov flux routine of Figure 5, the first instruction after the variables declarations (rhoL=VcL(VC_RHO)) is a DDT copy which is allowed since the assignment operator has been overloaded. This operation copies the components v and dv of $V_{c_L}(VC\_Rho)$ into the components v and dv of the variable $\rho_L$. The next line extracts the left-face side value of the velocity $u_L$ from the conservative vector. Overloading the division operator gives the following value for the derivative of $u_L$ with respect to $\rho_L$:

$$\boxed{\begin{aligned}&u_L=\frac{V_{c_L}(VC\_RhoU)}{V_{c_L}(VC\_Rho)}\Rightarrow u_L\%dv(1,1)=\frac{V_{c_L}(VC\_RhoU)\%dv(1,1)\times V_{c_L}(VC\_Rho)\%v - V_{c_L}(VC\_RhoU)\%v\times V_{c_L}(VC\_Rho)\%dv(1,1)}{\left(V_{c_L}(VC\_Rho)\%v\right)^2}\\ &\Rightarrow u_L\%dv(1,1)=\frac{0\times\rho_L-\rho_L u_L\times 1}{(\rho_L)^2}=\frac{-u_L}{\rho_L}\end{aligned}}$$

The same result is obtained by hand calculations. Denoting $m = \rho u$, the velocity is obtained as $u = \frac{m}{\rho}$. Hence,

$$\frac{\partial u_L}{\partial \rho_L} = \frac{\partial\left(\frac{m_L}{\rho_L}\right)}{\partial \rho_L} = -\frac{m_L}{\rho_L^2} = -\frac{\rho_L u_L}{\rho_L^2} = -\frac{u_L}{\rho_L}$$

Proceeding the same way for the next lines and assuming $S_{MAX} = u_L + c_L$, the following intermediate derivatives are obtained:

$$\boxed{\begin{aligned}&E_L\%dv(1,1)=\frac{-E_L}{\rho_L}\\ &p_L\%dv(1,1)=\frac{(\gamma-1)}{2}u_L^2\\ &c_L\%dv(1,1)=\frac{\gamma(\gamma-1)}{4\rho_L c_L}-\frac{c_L}{2\rho_L}\\ &S_{MAX}\%dv(1,1)=\frac{-u_L}{\rho_L}+\frac{\gamma(\gamma-1)}{4\rho_L c_L}-\frac{c_L}{2\rho_L}\end{aligned}}$$

Finally, the derivative of the Rusanov flux with respect to the left density is obtained as:

$$\frac{\partial F_{L,R}^*}{\partial \rho_L} = \frac{u_L + c_L}{2} - \left(\frac{\gamma(\gamma-1)u_L^2}{8\rho_L c_L} - \frac{u_L}{2\rho_L} - \frac{c_L}{4\rho_L}\right)(\rho_R - \rho_L)$$

which after some elementary algebraic manipulation gives the same expression as the one given in (7) obtained by hand differentiation.

## IV. Numerical examples

In this section numerical integration using time implicit schemes is performed on various flow models and spatial schemes to demonstrate the flexibility delivered by the automatic differentiation. One-dimensional and two-dimensional computations are considered using unstructured meshes decomposed in subdomains using Metis graph partitioning (Karypis & Kumar, 1998). Simulations are performed in parallel using the MPI protocol and the GNU-gfortran compiler. The linear systems arising from the implicit time discretization are solved by the PETSc libraries (Balay, et al., 2018) using block-Jacobi preconditioned GMRes solver (Saad & Schultz, 1986).

One-dimensional and two-dimensional Euler equations are addressed first on a shock-tube test problem followed by an unsteady transonic flow past a cylinder. Simulations of water-hammer using a two-phase flow model in equilibrium and a two-phase flow model in disequilibrium are carried out next.

1. **Euler equations**

    a. 1D Sod shock-tube test-case

The one-dimensional shock-tube test of Sod (1978) is performed first. The computational configuration is composed of a tube of 1m length with a membrane located at x=0.5m separating a chamber at atmospheric conditions on the left side and a low-pressure chamber on the right side. Air governed by the ideal gas equation of state is considered with polytropic coefficient $\gamma = 1.4$. The pressure is set to $p = 10^5 \text{Pa}$ to the left of the membrane and $p = 10^4 \text{Pa}$ to the right while the densities are set to $\rho = 1 \text{kg.m}^{-3}$ and $\rho = 0.125 \text{kg.m}^{-3}$ respectively. The computational domain is discretized with 10000 uniform cells. First-order finite volume computations are carried out using explicit Euler integration scheme with a time-step controlled by the CFL stability constraint $\text{CFL} = \lambda \frac{\Delta t}{\Delta x} = 0.5$, $\lambda$ being the fastest wave velocity over the domain at a given time $\lambda = \text{MAX}_\Omega \left( |u^n| + c^n \right)$. The time explicit numerical solution is compared to BDF1 with a single Newton iteration for increasing CFL numbers: 10,50 and 100.

Four numerical flux functions are tested in the present study: the approximate two-waves Riemann solver of Rusanov (Rusanov, 1962), the flux splitting scheme AUSM+ (Liou, 1996), the three-wave approximate Riemann solver HLLC (Toro, et al., 1994) and the Godunov scheme based on the exact Riemann solution (Godunov, 1959). The matrix Jacobian of all schemes is built using the ADOO method presented in Section III.2. While the algorithmic complexity of the Rusanov and HLLC schemes allow a relatively easy symbolic differentiation (see Rinaldi, et al., (2014) for the implicit HLLC scheme), the AUSM+ (see Colonia, et al., (2014) for analytic derivatives ) and the Godunov scheme require a fairly amount of work. The AUSM+ scheme involves numerous conditional branches for the evaluation of the split Mach number and split pressure.

The Godunov scheme is based on the exact solution of the Riemann problem (Godunov, 1959) and involves a root-finding algorithm for the determination of the star pressure, where symbolic differentiation often requires some assumptions at this stage. In contrast, ADOO is able to differentiate the entire numerical scheme, propagating the derivatives inside the fixed-point iterative algorithm in a fully transparent way. The only precaution to be made in the automatic differentiation of root-finding algorithms of Newton type used in the Riemann solver is to set a tolerance allowing both function and function derivative to be converged, the latter generally having a slower rate.

Density plots comparing explicit Euler and BDF1 time integration schemes are presented in Figure 6, for the Rusanov, HLLC, AUSM+ and Godunov flux functions. Excellent stability for all schemes is obtained even for relatively high CFL number. Unconditional SSP property in the linear case for BDF1 together with exact Jacobians transpose very well in this non-linear case.

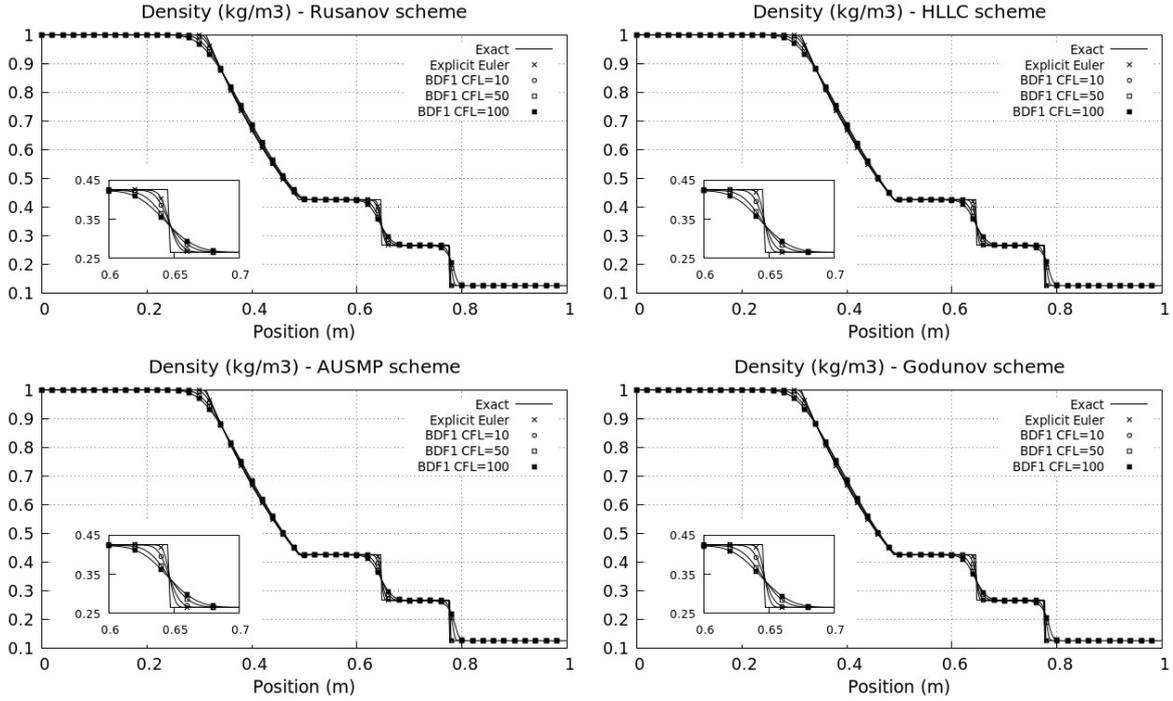

**Figure 6: 1D Sod shock-tube problem, density plot. Comparison of explicit Euler and BDF1 numerical solutions using the Rusanov, HLLC, AUSMP and Godunov (left to right, top to bottom respectively) schemes with various CFL numbers.**

BDF1 with HLLC scheme and CFL=100 results in a gain in computational time of about a factor 10. Note that in the present case, these 1D simulations have been carried out in the unstructured multi-dimensional code DALPHADT using the general GMRes linear solver provided by PETSc libraries. As such, the computational gain is representative of higher dimensions simulations.

ADOO time overhead has been measured with respect to analytic differentiation using the Rusanov flux function on a purely structured grid arrangement as well. Total simulation time has been recorded using both methods and use of ADOO showed negligible impact. Indeed, a time overhead lower than 0.1% with the fast block-tridiagonal Thomas algorithm has been observed. In multi-dimensional simulations with sparse non-symmetric matrices and general linear solvers, this overhead is expected to become even smaller.

b.  2D transonic flow past a circular cylinder

The second test-case using the compressible Euler equations is a 2D transonic flow around a circular cylinder of 1m diameter. The computational domain is rectangular (see Figure 7) with non-reflective boundary conditions (BC) assuming low-amplitude waves through algebraic acoustic relations. The circular cylinder BC assumes slip condition, consistent with the inviscid Euler equations. Free-stream BC as well as initial condition (IC) uses a Mach number (Ma) of 0.5 and an angle of attack of 0°. The domain is discretized with 50000 triangles.

A second-order finite volume spatial scheme is used. Cell-center gradients are computed using weighted least-squares based on a stencil composed of all cells sharing a vertex with the control volume. Multi-dimensional limiting process is ensured using a vertex-based extension (Park, et al., 2010) of the Barth and Jespersen limiter (Barth & Jespersen, 1989). The implicit operator is built using ADOO assuming dependence on direct neighbors only, e.g. assuming the set $\left\{\left\{\overline{\mathbf{Q}}_k\right\}\right\}_i$ composed of cells sharing a face with cell I (often referred as first-order Jacobian in the literature). Second-order spatial accuracy is ensured through the right-hand-side of the linear system where fluxes are computed from reconstructed states. This assumption greatly reduces the memory space needed for the implicit operator as well as increases the convergence rate of the linear solver. Indeed, on a regular 2D grid composed of equilateral triangles $\text{card}\left(\left\{\left\{\overline{\mathbf{Q}}_k\right\}\right\}_i\right) = 22$ using weighted least-squares gradient approximation together with vertex-based limiter of (Park, et al., 2010), resulting in 22 non-zero block matrices per row. Restriction of $\left\{\left\{\overline{\mathbf{Q}}_k\right\}\right\}_i$ to direct neighbors only involves a Jacobian matrix composed of 4 non-zero blocks per row. This assumption nevertheless introduces a slight inconsistency between the Jacobian matrix and the right-hand-side resulting in the loss of quadratic convergence of the non-linear residual.

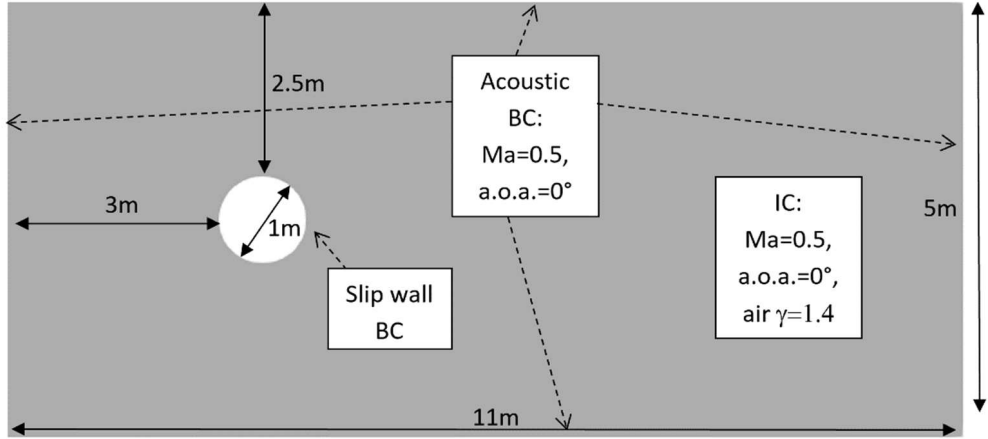

Figure 7: Single-phase transonic flow past a circular cylinder. Computational set-up.

Under these operating conditions, the flow becomes transonic near the top of the cylinder and attached shock-waves form downstream, ahead of the rear stagnation point. Due to the important total pressure loss across the shock and the high curvature, vorticity is generated, and the flow detaches from the wall resulting in an unsteady periodic flow (Salas, 1983) (see solution snapshots in Figure 8).

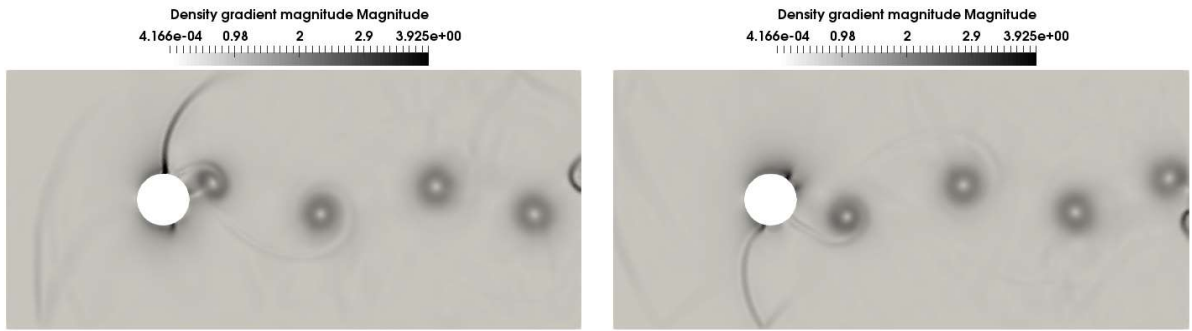

Figure 8: Single-phase transonic flow past a circular cylinder. Contours of density gradients magnitude when periodic regime is reached. Solutions are displayed with a time interval of 0.01s when periodic regime is reached.

Explicit and implicit computations have been carried out using the Rusanov, AUSM[+] and HLLC schemes for a physical time of 0.3s. Ideal gas equation of state is used with a polytropic coefficient $\gamma = 1.4$. The explicit scheme employs a second-order two-stage SSP-Explicit-Runge-Kutta method (SSPERGK-2) (Gottlieb, et al., 2001).

The lift coefficient is computed as $C_L = \dfrac{\int_s p\vec{n}\cdot\vec{e}_y dS}{\frac{1}{2}\rho_\infty u_\infty^2 S_w}$, with $\rho_\infty, u_\infty$ the free-stream density and velocity and $S_w = 1\text{m}^2$ the projected wetted surface. Lift coefficient is plotted as a function of time using AUSM[+] approximate Riemann solver, SSPERGK-2 with CFL=0.5 and one-iteration BDF1 with CFL=10 and CFL=20 in Figure 9 (left).

The results agree with those of Pandolfi & Larocca, (1989). The lift coefficient values obtained with the first-order implicit scheme are very close to those given by the second-order explicit scheme despite a slight phase lag increasing with the numerical integration time-step. This phase lag can be highly decreased using the second-order BDF2 scheme at the cost of solving the non-linear problem up to a reasonable tolerance, see Figure 9 (right). The Newton-BDF2 scheme with second-order spatial reconstruction is in fact a quasi-Newton approach due to the first-order Jacobian assumption, and only linear convergence has been obtained. A decrease factor of $10^6$ has been imposed to the magnitude of the non-linear residual with a maximum of 10 Newton iterations. Newton-BDF2 is however a rather costly method and no computational gain has been observed with respect to the explicit integration. On the other hand, a factor of 8 in computational time has been obtained using BDF1 (CFL=20) with respect to SSPERGK-2 on a 16-cores Intel Xeon E5-2687W-v2.

The simulation has been carried out using the Rusanov and HLLC fluxes and results are displayed in Figure 10. Explicit and implicit results are in overall good agreement.

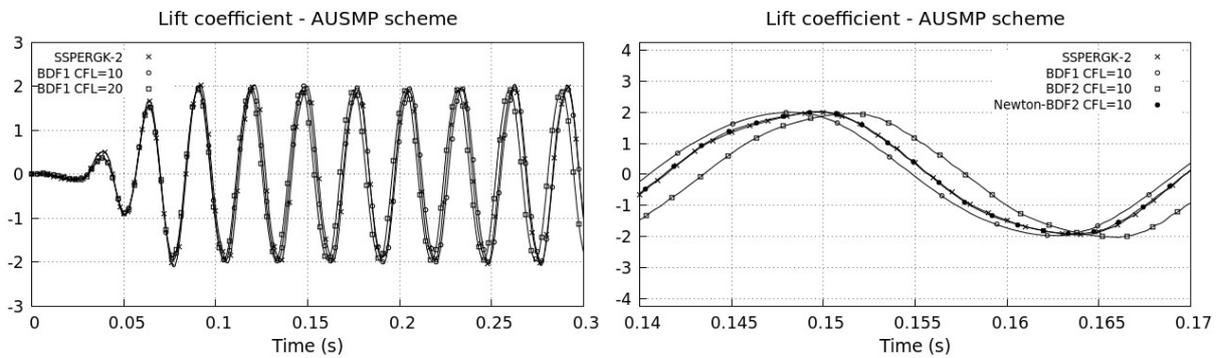

Figure 9: 2D transonic flow past a circular cylinder. left: comparison of lift coefficient using explicit and implicit schemes based on AUSM+ spatial flux and right: BDF2 and Newton-BDF2 schemes compared to BDF1 and SSPERGK-2.

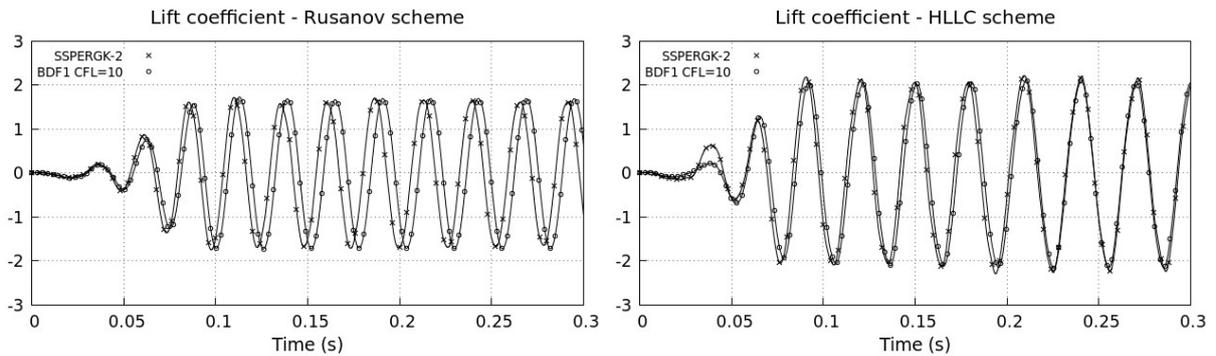

Figure 10: 2D transonic flow past a circular cylinder. Comparison of lift coefficient using explicit and implicit schemes based on Rusanov and HLLC spatial fluxes (left and right respectively).

All the tested implicit schemes are in good agreement with the explicit solution and results from the literature.

ADOO based implicit schemes are now considered for more sophisticated flow models for which derivation of numerical fluxes is a challenging task.

2. **Two-phase flow models in full equilibrium and full disequilibrium**

Two compressible two-phase flow models are presented in this section. The implicit discretization using ADOO is applied and verified on the numerical simulation of water-hammer two phase flow.

a. Parent model in full disequilibrium

The symmetric variant (Saurel , et al., 2003) of the compressible two-phase flow model of (Baer & Nunziato, 1986) in the absence of mass and heat transfer reads in the following 1D compact form:

$$\frac{\partial \mathbf{Q}}{\partial t} + \frac{\partial \mathbf{F}(\mathbf{Q})}{\partial x} + \mathbf{G}(\mathbf{Q})\frac{\partial \mathbf{H}(\mathbf{Q})}{\partial x} = \mathbf{S}(\mathbf{Q}) \qquad (8)$$

with

$$\mathbf{Q} = \begin{pmatrix} \alpha_1 \\ \alpha_1\rho_1 \\ \alpha_1\rho_1 u_1 \\ \alpha_1\rho_1 E_1 \\ \alpha_2\rho_2 \\ \alpha_2\rho_2 u_2 \\ \alpha_2\rho_2 E_2 \end{pmatrix}, \mathbf{F}(\mathbf{Q}) = \begin{pmatrix} 0 \\ \alpha_1\rho_1 u_1 \\ \alpha_1(\rho_1 u_1^2 + p_1) \\ \alpha_1 u_1(\rho_1 E_1 + p_1) \\ \alpha_2\rho_2 u_2 \\ \alpha_2(\rho_2 u_2^2 + p_2) \\ \alpha_2 u_2(\rho_2 E_2 + p_2) \end{pmatrix}, \mathbf{G}(\mathbf{Q}) = \begin{pmatrix} u_I \\ 0 \\ -p_I \\ -p_I u_I \\ 0 \\ -p_I \\ -p_I u_I \end{pmatrix}, \mathbf{H}(\mathbf{Q}) = \begin{pmatrix} \alpha_1 \\ 0 \\ \alpha_1 \\ \alpha_1 \\ 0 \\ \alpha_2 \\ \alpha_2 \end{pmatrix} \qquad (9)$$

and

$$\mathbf{S}(\mathbf{Q}) = \begin{pmatrix} \mu(p_1 - p_2) \\ 0 \\ \lambda(u_2 - u_1) \\ \lambda\tilde{u}_I(u_2 - u_1) - \mu\tilde{p}_I(p_1 - p_2) \\ 0 \\ -\lambda(u_2 - u_1) \\ -\lambda\tilde{u}_I(u_2 - u_1) + \mu\tilde{p}_I(p_1 - p_2) \end{pmatrix} \qquad (10)$$

In this system, the subscripts 1 and 2 refer to phase 1 and 2 respectively. $\alpha_k, \rho_k, u_k, p_k, E_k, T_k$ ($k=1,2$) are the volume fraction, density, velocity, pressure, total energy and temperature of each phase. $u_I, p_I$ are the interfacial velocity and pressure modeled as in (Saurel , et al., 2003):

$$u_I = \frac{Z_1 u_1 + Z_2 u_2}{Z_1 + Z_2} + \mathrm{sgn}\left(\frac{\partial \alpha_1}{\partial x}\right)\frac{p_2 - p_1}{Z_1 + Z_2}, \quad p_I = \frac{Z_2 p_1 + Z_1 p_2}{Z_1 + Z_2} + \mathrm{sgn}\left(\frac{\partial \alpha_1}{\partial x}\right)\frac{Z_1 Z_2}{Z_1 + Z_2}(u_2 - u_1) \qquad (11)$$

where $Z_k = \rho_k c_k$ is the acoustic impedance of phase k with $c_k$ the associated sound speed. The homogeneous part of this model considers each phase as compressible, evolving with its own velocity, pressure and temperature. The source term vector $\mathbf{S}(\mathbf{Q})$ contains a closure law for pressure disequilibrium and drag effects, where $\mu$ controls the rate at which pressure equilibrium is reached and $\lambda$ is the coefficient of friction. The volume average pressure and interface velocity are given by:

$$\tilde{u}_I = \frac{Z_1 u_1 + Z_2 u_2}{Z_1 + Z_2}, \quad \tilde{p}_I = \frac{Z_2 p_1 + Z_1 p_2}{Z_1 + Z_2} \qquad (12)$$

The system is closed by an equation of state for each phase, stiffened-gas in this work:

$$E_k = e_k(\rho_k, p_k) + \frac{1}{2}u_k^2 = \frac{p_k + \gamma_k p_{SG,k}}{\rho_k(\gamma_k - 1)} + \frac{1}{2}u_k^2 \quad (13)$$

and by the saturation constraint:

$$\alpha_1 + \alpha_2 = 1 \quad (14)$$

The compressible two-phase flow model (8)-(14) is a non-conservative hyperbolic system. The intercell flux functions arising from the spatial finite volume discretization are solved using an algebraic HLLC-type approximate Riemann solver detailed in (Furfaro & Saurel, 2015) and summarized in Appendix B.

Two 1D shock tube problems are solved using explicit and implicit time integrators. The first problem is a water-air shock tube without relaxations, and the second test is a shock tube in a water-air mixture with instantaneous pressure relaxation and drag effects.

*Water-air shock tube without relaxations*

Initially a 1m length tube is filled with pure water on the left side of a membrane located at x=0.8m and pure air on the right side. Water pressure is set to 0.2GPa while air pressure is set to 0.1MPa. Densities of water and air are initially set to 1000kg/m³ and 1 kg/m³ respectively. Water phase is governed by the stiffened gas EOS with parameters $p_{SG,1} = 1\text{GPa}$ and $\gamma = 2.35$ and air is governed by the ideal gas law with $\gamma = 1.4$. A small volume fraction of air is present in the water and vice versa ($\alpha = 10^{-6}$). The mesh is composed of 2000 elements and first order spatial discretization is used. The simulation time is set to 276µs.

This test case in the absence of pressure and velocity relaxations is very challenging. Indeed, interface conditions across the two-phase contact of equal pressure and normal velocity are not trivial to achieve as they are solely satisfied by the consistent discretization of the non-conservative products present in the model. This problem is thus a strong benchmark for numerical schemes applied to the two-phase flow model out of equilibrium.

Explicit Euler temporal integration with CFL=0.5 and implicit methods with CFL=20 using Newton-BDF1, Newton-BDF2 and Newton-SSPSDIRK schemes have been compared. Single-step BDF methods failed on this test case for CFL numbers greater than 5. Newton method tolerance has been set to 10⁻⁶, reached in less than 10 iterations thanks to the Jacobian exactness yielding quadratic convergence.

Volume fraction of water is displayed in Figure 11. Explicit and implicit schemes are undistinguishable, and they all yield a perfectly bounded solution.

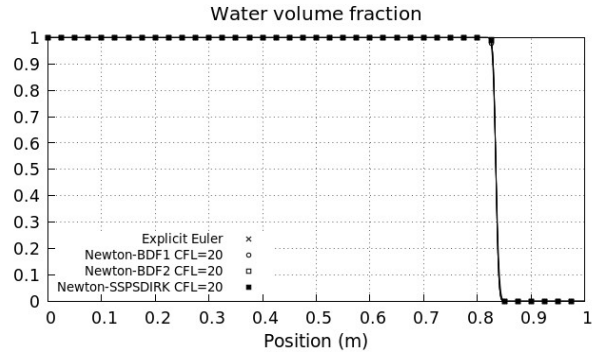

**Figure 11: Water-air shock tube without relaxations. Volume fraction of water using explicit and implicit schemes.**

Mixture pressure and velocity as well as pressures and velocities of both phases are plot in Figure 12. It can be seen on the top graphs that BDF1 is more diffusive than the explicit Euler integration but is strongly stable. Newton-BDF2 and Newton-SSPSDIRK which are second order accurate give very similar solutions matching the explicit solution. A small oscillation is however present at the tail of the rarefaction wave using Newton-BDF2 scheme where Newton-SSPSDIRK remains stable. Second order implicit methods are only conditionally SSP, SSPSDIRK with a less stringent time-step limit.

Bottom graphs of Figure 12 show both phases pressures and velocities. Only Newton-SSPSDIRK results are shown for the sake of conciseness, explicit solution can be found in (Furfaro & Saurel, 2015). Interface conditions are fulfilled: at the contact pressures and velocities of both phases match perfectly.

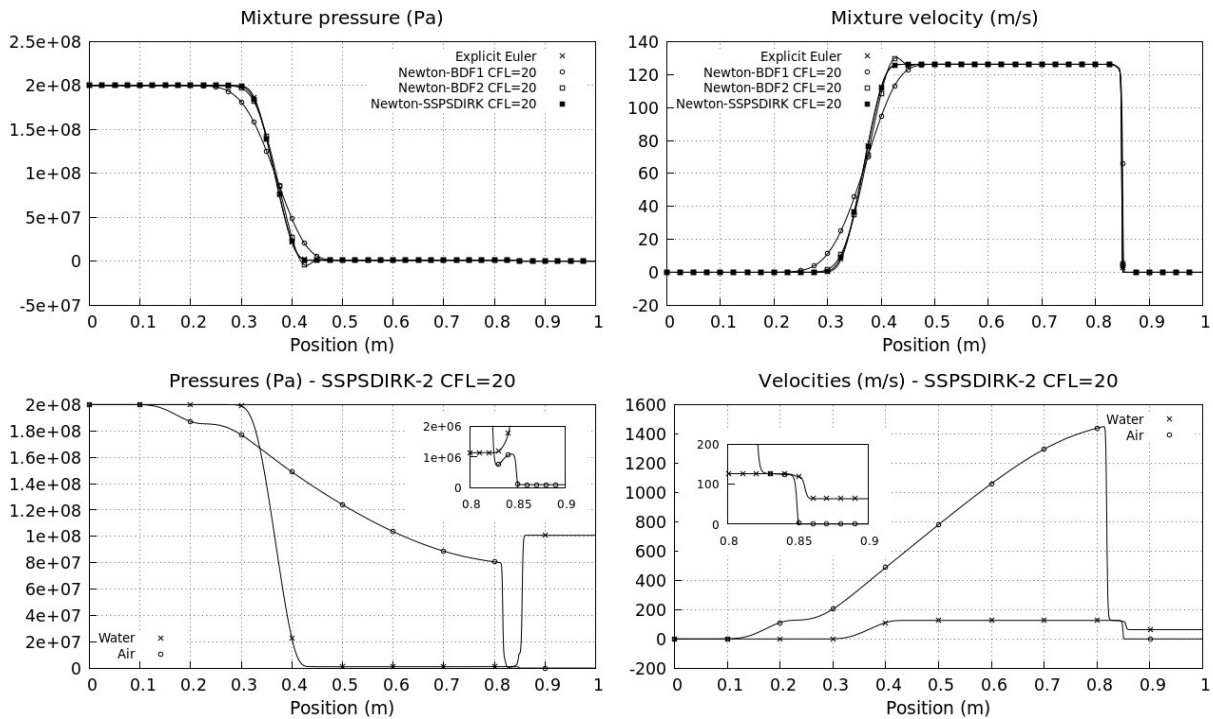

**Figure 12: Water-air shock tube without relaxations. Comparison between explicit and implicit time integrators. Top: mixture pressure and mixture velocity. Bottom: pressures and velocities of both phases showing correct fulfillment of interface conditions.**

*Water-air mixture shock tube with pressure relaxation and drag effects*

The second test case considered with the two-phase model out of equilibrium is a shock tube composed of a water-air mixture. A 1m length tube is initially filled with water and air in proportion $\alpha = 0.1$ and $\alpha = 0.9$. A membrane is placed at $x = 0.5\mathrm{m}$ separating a hot pressure chamber at 100bar to the left and 1b to the right.

Densities of water and air are set to 1000kg/m³ and 1kg/m³ respectively. Water phase is governed by the stiffened gas EOS with parameters $p_{SG,1} = 1\text{GPa}$ and $\gamma = 2.35$ and air is governed by the ideal gas law with $\gamma = 1.4$. The mesh is composed of 2000 elements and first order spatial discretization is used. The simulation time is set to 790µs.

Instantaneous pressure relaxation is applied following a time splitting procedure detailed in (Furfaro & Saurel, 2015). Drag effects are accounted for considering constant water bubble radius $R_w = 5\text{mm}$. The exchange interfacial area per unit volume ($A_I$) dependence is thus simplified. It is assumed dependent only on water volume fraction ($\alpha_w$) evolution which in turn is driven by pressure equilibrium. The total friction coefficient $\lambda$ is then modelled as:

$$\lambda = \frac{Z_1 Z_2}{Z_1 + Z_2} A_I = \frac{Z_1 Z_2}{Z_1 + Z_2} \frac{3\alpha_w}{R_w}$$

This finite rate momentum exchange is integrated following a first-order explicit time-splitting method.

A comparison between explicit Euler temporal integration using CFL=0.5 and Newton-SSPSDIRK scheme using CFL=10,20 and 30 is carried out. Pressure is displayed in Figure 13, showing overall good agreement between the explicit and the implicit solutions. The dominant error is due to the time-splitting strategy. Indeed, the difference between the explicit solution and SSPSDIRK with CFL=10 is much higher than the differences between the various implicit solutions using increasing CFL.

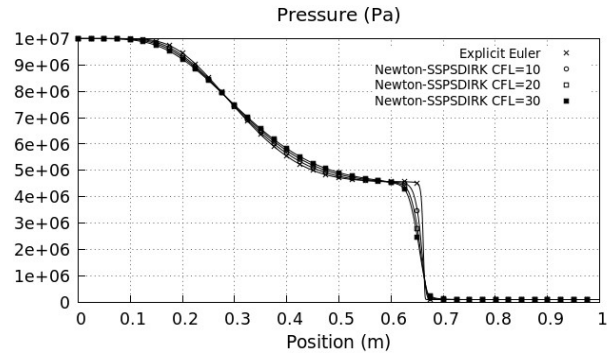

**Figure 13: Water-air mixture shock tube with pressure relaxation and drag effects. Relaxed pressure using explicit and implicit time integrators.**

Water and air velocities are displayed in Figure 14. Momentum exchanges driven by drag effects are quite stiff under these operating conditions as shown by the relatively small velocity drift. Overall good agreement between the explicit and the implicit solutions is visible.

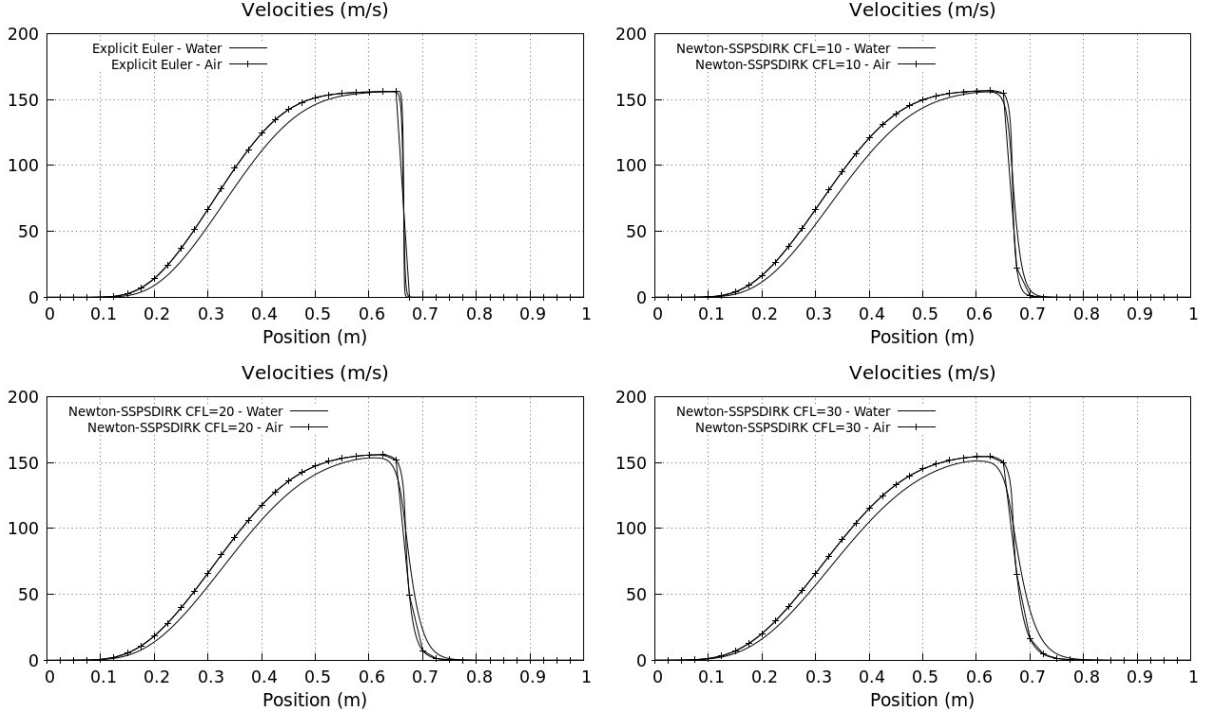

Figure 14: Water-air mixture shock tube with pressure relaxation and drag effects. Water and air velocities using explicit and implicit time integrators.

b. Reduced model in full equilibrium

Another two-phase flow model is considered as well for numerical experiments. An asymptotic analysis in the stiff limit of velocity, pressure and temperature of system (8)-(14) yields the following two-phase model in full equilibrium (Le Martelot, et al., 2014) (Chiapolino, et al., 2016):

$$\frac{\partial \mathbf{Q}}{\partial t} + \frac{\partial \mathbf{F}(\mathbf{Q})}{\partial x} = 0, \mathbf{Q} = \begin{pmatrix} \rho \\ \rho u \\ \rho E \\ \rho Y_1 \end{pmatrix}, \mathbf{F}(\mathbf{Q}) = \begin{pmatrix} \rho u \\ \rho u^2 + p \\ u(\rho E + p) \\ \rho u Y_1 \end{pmatrix} \quad (15)$$

where $\rho, u, p, E, Y_1$ denote the mixture density, velocity, pressure, total energy and mass fraction of phase 1 respectively. The system is closed by the saturation constraint:

$$Y_1 + Y_2 = 1 \quad (16)$$

and by a mixture equation of state satisfying the following principles of conservation of mixture internal energy and mixture specific volume:

$$e = Y_1 e_1(p, T) + Y_2 e_2(p, T)$$
$$v = Y_1 v_1(p, T) + Y_2 v_2(p, T) \quad (17)$$

Considering stiffened-gas equation of state for phase 1 and ideal gas equation of state for phase 2, system (17) admits a unique physical solution for the mixture pressure:

$$p = \frac{1}{2}(A_1 + A_2 - p_{SG,1}) + \sqrt{\frac{1}{4}(A_2 - A_1 + p_{SG,1})^2 + A_1 A_2}$$

$$\text{with } A_1 = (\rho e - p_{SG,1})\frac{Y_1(\gamma_1 - 1)c_{v_1}}{Y_1 c_{v_1} + Y_2 c_{v_2}}, A_2 = \rho e \frac{Y_2(\gamma_2 - 1)c_{v_2}}{Y_1 c_{v_1} + Y_2 c_{v_2}} \quad (18)$$

Equations (15),(16) and (18) form a conservative hyperbolic system. The sound speed associated to this system of equations is very well approximated by the simple Wood formula (Wood, 1930). The intercell flux function arising from the spatial finite volume discretization is solved using an algebraic HLLC-type approximate Riemann solver (Saurel, et al., 2006).

Two 1D problems are solved using explicit and implicit time integrators. The first test is a shock tube in a water-air mixture while the second problem is a double rarefaction wave. Multi-dimensional extension is addressed through a sonic jet problem and a gas-gas single mode Rayleigh-Taylor instability

*Water-air mixture shock tube*

A 1m length tube is filled with a mixture of water and air. Water mass fraction is set to 2%. Initially, a membrane located at x=0.5m separates a high-pressure chamber to a low-pressure one. On the left side, pressure is set to 0.2MPa while it is set to 0.1MPa on the right side. Temperature is imposed to 293K in all the domain. Water phase is governed by the stiffened gas EOS with parameters $p_{SG,1} = 1\text{GPa}$ and $\gamma = 2.35$ and air is governed by the ideal gas law with $\gamma = 1.4$. The mesh is composed of 10000 elements and first order spatial discretization is used. The simulation time is set to 1ms.

Explicit and implicit integration is carried out. The explicit scheme uses first order Euler integration with CFL number set to 0.5, while the implicit scheme uses single-step BDF1 with CFL=10,20 and 40. Mixture density is displayed at the final time in Figure 15. Under these mild conditions single-step BDF1 remains stable even for a CFL number of 40, yielding a computational gain of a factor 10 with respect to explicit integration.

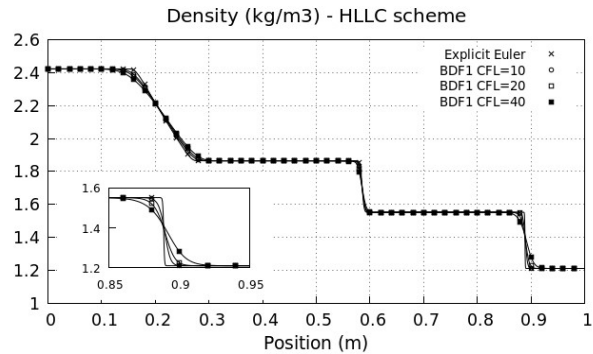

Figure 15: Water-air mixture shock tube. Mixture density at time t=1ms using explicit Euler and implicit BDF1 schemes.

*Double rarefaction wave*

The second 1D test case considered using the two-phase model in mechanical and thermal equilibrium is a double rarefaction wave problem. This test is particularly interesting to benchmark numerical schemes as vacuum conditions are reached at the center.

A 1m length tube is considered filled with almost pure water (air mass fraction equal to $10^{-6}$) at atmospheric conditions p=0.1MPa and T=293K. Initially a membrane is located at x=0.5m. Velocity is set to -10m/s and 10m/s to the left and to the right of the membrane respectively. The domain is discretized into 10000 elements and the simulation time is set to 1.5ms. Explicit first order Euler integration with CFL number equal to 0.5 and implicit single-step BDF1 scheme with CFL=10,20 and 40 are used with a first order spatial discretization.

Pressure and velocity at time t=1.5ms are displayed in Figure 16. Single-step BDF1 scheme remain stable in all cases without pressure positivity violation. At a CFL of 40, the gain in CPU time is about a factor 10 with respect to explicit Euler integration.

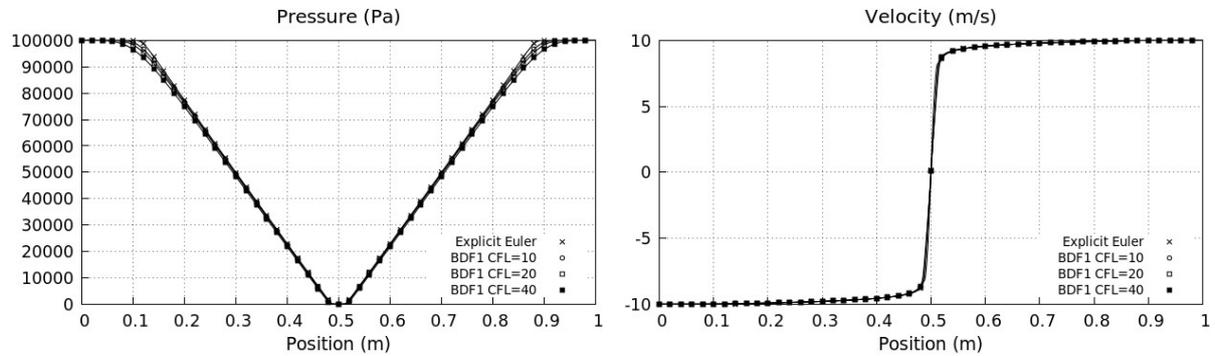

Figure 16: Double rarefaction wave. Velocity (left) and pressure (right) using explicit Euler and implicit BDF1 schemes.

Multi-dimensional tests are now addressed. A 2D two-phase sonic jet simulation is carried out first, then a 2D two-phase single mode Rayleigh-Taylor instability is computed.

*2D sonic jet*

A 2D jet configuration is set up as described in Figure 17. Initially the domain is composed of a water-air mixture in equal mass proportion at atmospheric conditions. Water phase is governed by the stiffened gas EOS with parameters $p_{SG,l} = 1\text{GPa}$ and $\gamma = 2.35$ and air is governed by the ideal gas law with $\gamma = 1.4$. At the left side, a two-phase pressure tank-inlet boundary condition is imposed (see Appendix A). Tank pressure is set to 0.2MPa yielding a chocked flow through a convergent-divergent nozzle and a sonic jet develops.

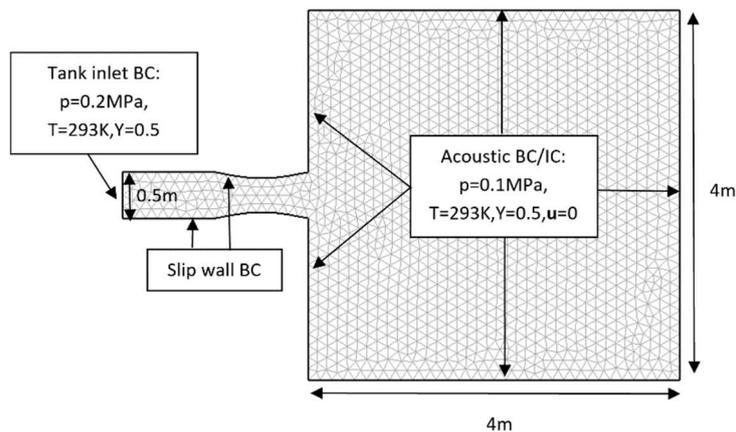

Figure 17: 2D sonic jet computational set-up.

The domain is discretized using an unstructured mesh composed of about 250000 triangles (the mesh edges length is multiplied by a factor 10 in Figure 17). Second order spatial discretization is applied using Barth and Jespersen limiter and the simulation time is set to 21ms. Explicit temporal integration uses second order SSPERGK scheme with CFL number equal to 0.5. Implicit schemes use single-step first order BDF and second order Newton-SSPSDIRK both using CFL equal to 20. In case of Newton-SSPSDIRK scheme, as quadratic convergence cannot be achieved due to the first order Jacobian simplification, a limit of five iterations to reach a non-linear residual magnitude of $10^{-6}$ has been imposed.

Two-phase tank-inlet condition implies the solution to a non-linear equation to approximate the boundary flux. The Jacobian boundary contribution is computed in a straightforward manner using ADOO as detailed in Appendix A.

Contours of density at time 14ms and 21ms for the three temporal schemes are displayed in Figure 18. No significant diffusion is observed on the jet using BDF1 scheme, in contrast the shock wave is slightly smeared. At this CFL number, the higher order Newton-SSPSDIRK scheme predicts a sharper shock front but some non-amplifying oscillations are present.

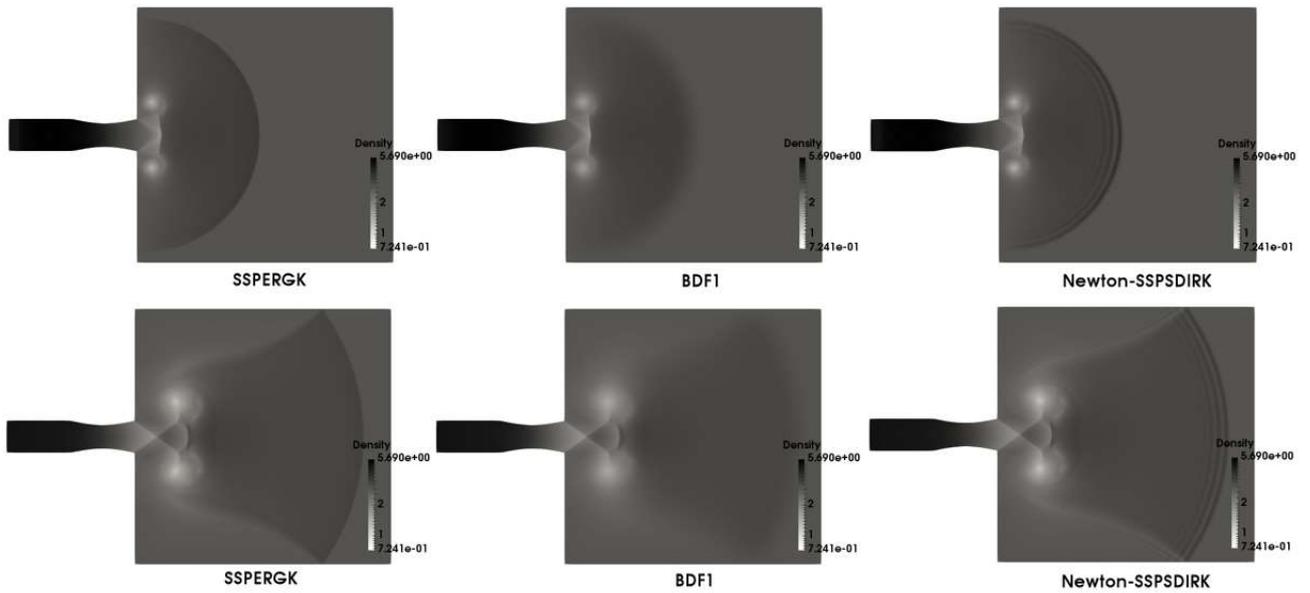

Figure 18: 2D jet at time t=14ms (top) and t=21ms (bottom). Mixture density contours.

Newton-SSPSDIRK scheme with an imposed maximum of five non-linear iterations per stage requires about 10 linear systems to be solved at each time step. At a CFL number of 20, the CPU gain with respect to two-stage explicit SSPERGK scheme is about 17%. Single-step BDF scheme which only requires one linear system solution per time step gives a CPU gain factor of about 5.

*Air-Helium Rayleigh-Taylor instability*

A 2D single-mode Rayleigh-Taylor configuration is set up as described in Figure 19. Initially the domain is composed of a pure light gas (helium) on the bottom and a denser gas on the top (air). An initial perturbed interface separates the two gases to trigger a single-mode instability driven by gravity effects. Gravity magnitude has been set to 1000m/$s^2$. The computational domain is all bounded by slip wall boundary conditions. Air and helium are governed by the ideal gas law with polytropic coefficients 1.4 and 1.67 respectively.

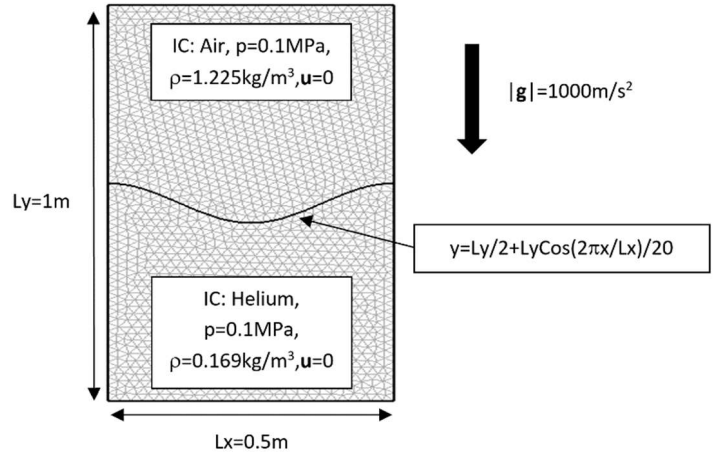

**Figure 19: 2D single-mode Rayleigh-Taylor instability computational set up.**

Final simulation time is set to 37.5ms. In the absence of viscous and surface tension effects, the interface is highly unstable and subject to chaotic behavior which can be triggered by small perturbations induced for example using distinct spatial numerical discretization (Liska & Wendroff, 2004). In our context, the spatial discretization is fixed using second-order HLLC scheme on a mesh composed of about 300000 triangles.

Explicit temporal integration uses second order SSPERGK scheme with CFL number equal to 0.5. Implicit schemes use single-step first order BDF and second order Newton-SSPSDIRK both using CFL equal to 100. In case of Newton-SSPSDIRK scheme, a limit of five iterations to reach a non-linear residual magnitude of $10^{-6}$ has been imposed. Contours of mixture density are shown for the three temporal schemes at two time intervals in Figure 20.

Single-step first order BDF scheme is more diffusive than the explicit scheme resulting in a slightly altered mushroom shape. In contrast, second order implicit SSPSDIRK scheme gives a remarkably similar solution as the explicit one.

The CPU speed-up using implicit SSPSDIRK scheme is about a factor 4 while it reaches a factor of about 27 using single-step first order BDF method.

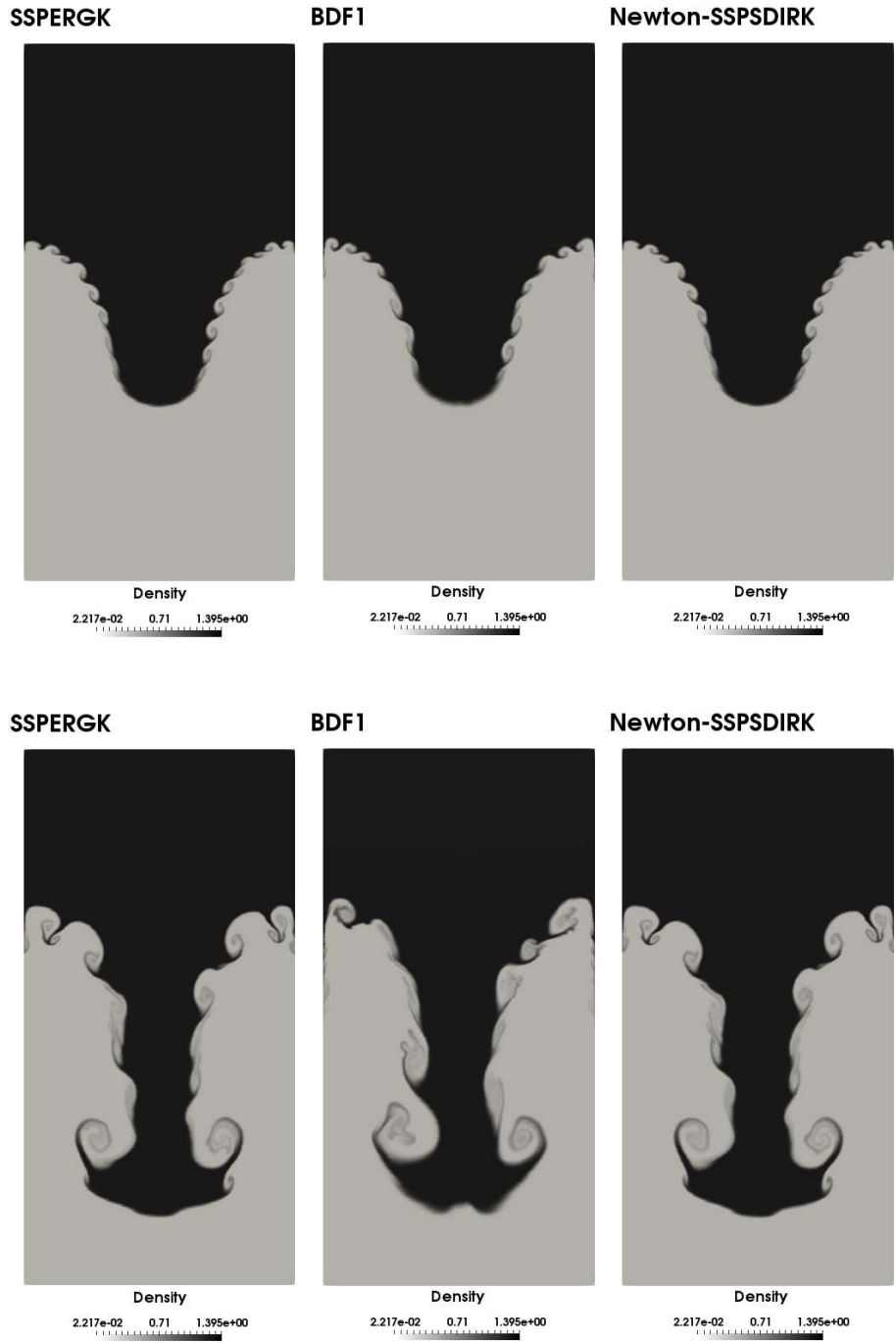

Figure 20: Single mode Rayleigh-Taylor instability at time t=22.5ms (top) and t=37.5ms (bottom). Mixture density contours using explicit and implicit time integrators.

## V. Conclusion

A numerical procedure to evaluate the matrix Jacobian coming from the implicit discretization of hyperbolic systems has been presented. The approach is based on the forward mode of ADOO and showed numerous advantages:

- It is easily applicable to existing codes with languages compilers supporting user-defined DDT and operator overloading;

- It is highly flexible with models and numerical schemes adjustments as the user do not need to worry about the implementation of the derivatives;
- It handles complex numerical algorithms such as root-finding methods or conditional branches differentiation in a straightforward manner,
- The overhead associated to operations on DDTs is negligible in the present context as most of the computational time is spent in the linear solver.

The ADOO method has been applied to the implicit discretization of various flow models, involving root-finding methods and complex equations of state. In the test-cases needing unsteady residual resolution, the accurate Jacobian evaluation allowed very fast convergence of Newton's method (always quadratic for first-order in space discretization).

ADOO allowed the implicit discretization of the compressible two-phase flow model of Baer and Nunziato. To the author's knowledge, the present work corresponds to the first successful attempt with a fully implicit approach.

## Appendix A: Tank-inlet boundary condition

The tank-inlet boundary condition is obtained though the solution of a semi-Riemann problem described in Figure 21.

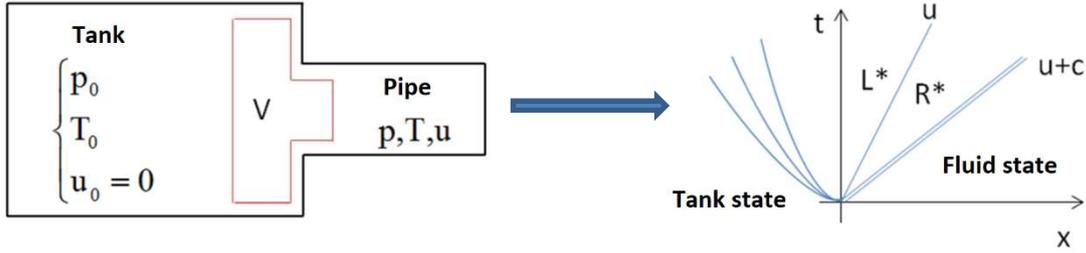

Figure 21: Tank-inlet boundary condition description.

Assuming a stationary flow between the tank and the inlet, the integration of the energy equation associated to the Euler equations on volume V yields:

$$\int_V \vec{\nabla}.(\rho H \vec{u}) dV = 0 \Leftrightarrow (\rho u H S)^*_L - (\rho u H S)_0 = 0 \overset{\text{mass cons}°}{\Rightarrow} H^*_L = H_0$$

which gives a first relation linking the tank state to the left Riemann state.

The enthalpy is conserved across the left curved wave. Considering an isentropic flow in the tank, the speed of sound $c^2 = \left(\frac{\partial p}{\partial \rho}\right)_s$ is assumed constant across the left wave: $c^{*2}_L = c_0^2 \approx \frac{p^*_L - p_0}{\rho^*_L - \rho_0}$. This gives a second relation linking the tank state to the left Riemann state: $\rho^*_L = \rho_0 + \frac{p^*_L - p_0}{c_0^2}$. Acoustic relations are assumed through the right wave:

$$p^*_R - \rho_R c_R u^*_R = p_R - \rho_R c_R u_R$$
$$\Rightarrow p^*_R(u^*_R) = p_R + \rho_R c_R (u^*_R - u_R)$$

The middle wave being necessarily a contact, equality of pressure and normal velocity is imposed yielding $p^*_L = p^*_R, u^*_L = u^*_R$. Putting all relations together yields to:

$$\left.\begin{aligned}
&e(p^*_L, \rho^*_L) + \frac{p^*_L}{\rho^*_L} + \frac{1}{2}u^{*2}_L - H_0 = 0 \\
&\rho^*_L = f(p^*_L) = \rho_0 + \frac{p^*_L - p_0}{c_0^2} \\
&u^*_R = g(p^*_R) = \frac{p^*_R - p_R}{\rho_R c_R} + u_R \\
&p^*_L = p^*_R \\
&u^*_L = u^*_R
\end{aligned}\right\} \Rightarrow e(p^*, f(p^*)) + \frac{p^*}{f(p^*)} + \frac{1}{2}g(p^*)^2 - H_0 = h(p^*) = 0$$

After application of the equation of state, this non-linear relation is solved numerically using Newton's method. Classical sampling is then performed, and the flux is computed from the Riemann solution. ADOO is propagated to the Newton's method in a transparent way, but with an additional convergence criterium based on the magnitude of the derivative of function $h(p^*)$.

Extension to the two-phase reduced model is straightforward considering the mixture equation of state. A prototype subroutine is given as example in Figure 22.

```fortran
SUBROUTINE TankInlet(p0,rho0,c0,Y0,H0,VcR,fluxbc)
REAL(KIND=RP),        INTENT(IN)    :: p0,rho0,c0,Y0,H0
TYPE(AutoDiffType),INTENT(INOUT) :: VcR(:)
TYPE(AutoDiffType),INTENT(OUT)   :: fluxbc(:)
INTEGER                          :: var,nit
TYPE(AutoDiffType)               :: rhor,ur,Etotr,Yr,er,pr,cr
TYPE(AutoDiffType)               :: pstar,rhostarl,rhostarr,rhostar,ustar,Ystar,estar,fstar,dfstar

! Derivatives initialization
DO var=1,Ncons
  VcR(var)%dv(:)=0.0_RP
  VcR(var)%dv(var)=1.0_RP
END DO

rhor=VcR(VC_Rho)
ur=VcR(VC_RhoU)/rhor
Etotr=VcR(VC_RhoE)/rhor
Yr=VcR(VC_RhoY)/rhor
er=Etotr-0.5_RP*ur**2
pr=prey(rhor,er,Yr)
cr=cpry(pr,rhor,Yr)

! Newton method for the solution of h(pstar)=0.
pstar=p0
DO nit=1,nitmax
  rhostarl = rho0+(pstar-p0)/c0**2
  ustar    = (pstar-pr)/(rhor*cr)+ur
  fstar    = FuncHP(p0,rho0,c0,Y0,rhor,pr,cr,ur,rhostarl,ustar,pstar)
  dfstar   = dFuncHPdp(p0,rho0,c0,Y0,rhor,pr,cr,ur,rhostarl,ustar,pstar)
  IF(fstar<eps.AND.dfstar<eps) EXIT
  pstar    = pstar-fstar/dfstar
END DO

IF(nit==nitmax+1) STOP "Newton convergence failed in TankInlet."

! Sampling
IF(ustar>0.0_RP)THEN
  Ystar   = Y0
  rhostar = rhostarl
ELSE
  Ystar   = Yr
  rhostarr = rhor+(pstar-pr)/cr**2
  rhostar = rhostarr
ENDIF

! Boundary flux. fluxbc(:)%dv contains derivative blocks w.r.t. VcR.
fluxbc(VC_Rho)  = rhostar*ustar
fluxbc(VC_RhoY) = rhostar*ustar*Ystar
fluxbc(VC_RhoU) = rhostar*ustar**2+pstar
fluxbc(VC_RhoE) = ustar*(rhostar*(epry(pstar,rhostar,Ystar)+0.5_RP*ustar**2)+pstar)

END SUBROUTINE TankInlet
```

Figure 22: FORTRAN routine for the computation of tank-inlet boundary flux. The resulting data structure fluxbc contains the derivative blocks of the flux with respect to the vector of conservative variables.

# Appendix B: HLLC solver for the two-phase model in disequilibrium

The two-phase model in disequilibrium can be rewritten under the following compact form for a generic phase k:

$$\frac{\partial \mathbf{Q}_k}{\partial t} + \frac{\partial \mathbf{F}_k(\mathbf{Q}_k)}{\partial x} + \alpha_k \frac{\partial \mathbf{H}_k(\mathbf{Q}_k)}{\partial x} = 0, \mathbf{Q}_k = \begin{pmatrix} \alpha_k \\ (\alpha\rho)_k \\ (\alpha\rho u)_k \\ (\alpha\rho E)_k \end{pmatrix}, \mathbf{F}_k(\mathbf{Q}_k) = \begin{pmatrix} \alpha_k u_I \\ (\alpha\rho u)_k \\ \alpha_k(\rho u^2 + p)_k - \alpha_k p_I \\ \alpha_k(\rho E + p)_k u_k - \alpha_k p_I u_I \end{pmatrix}, \mathbf{H}_k(\mathbf{Q}_k) = \begin{pmatrix} -u_I \\ 0 \\ p_I \\ p_I u_I \end{pmatrix} \quad (19)$$

The first step consists in approximating the interfacial terms $u_I$ et $p_I$. This step takes its roots in the Discrete Equations Method (Saurel, et al., 2003), given a 2D topology composed of independant channels as illustrated in Figure 23.

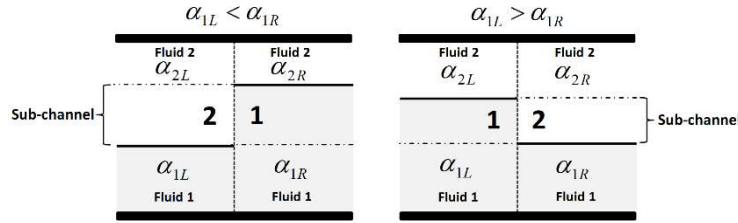

**Figure 23: 2D topology showing two types of contacts at a given element boundary.**

On a given cell boundary, three types of contacts are admissible : two mono-phasic contacs 1-1 and 2-2 and one contact 2-1 if the phase 1 volume fraction is higher on the right side or 1-2 if it is lower. Interfacial pressure and velocity are approximated as the solution of the Riemann problem associated to the two-fluid Euler equations. As the Riemann solution depends only on the left and on the right state, $u_I$ and $p_I$ become locally constant during a time-step.

Focusing on a 2-1 contact, the wave pattern is shown in Figure 24.

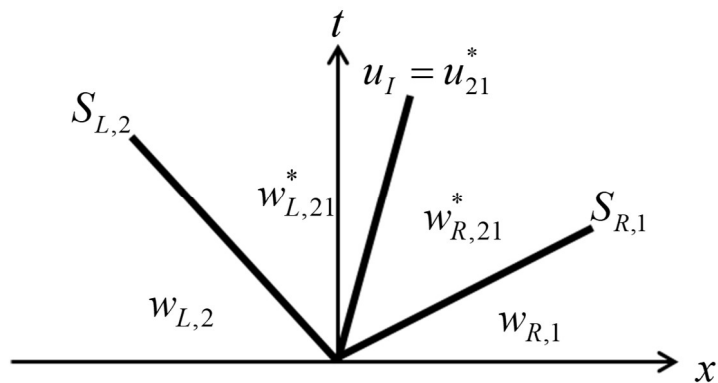

**Figure 24: Riemann wave pattern associated to contact 2-1.**

Wave speeds $S_{L,2}, S_{R,1}$ are computed following Davis approximation (Davis, 1988):

$$S_{R,1} = \text{Max}\left(u_{L,1} + c_{L,1}, u_{R,1} + c_{R,1}\right)$$
$$S_{L,2} = \text{Min}\left(u_{L,2} - c_{L,2}, u_{R,2} - c_{R,2}\right)$$

The interfacial variables $u_I$ and $p_I$ are computed HLL (Harten, et al., 1983) and HLLC (Toro, et al., 1994) approximations respectively:

$$u_I = u_{21}^{*,HLL} = \frac{\left(\rho u^2 + p\right)_{R,1} - \left(\rho u^2 + p\right)_{L,2} + S_{L,2}\left(\rho u\right)_{L,2} - S_{R,1}\left(\rho u\right)_{R,1}}{\left(\rho u\right)_{R,1} - \left(\rho u\right)_{L,2} + S_{L,2}\rho_{L,2} - S_{R,1}\rho_{R,1}}$$

$$p_I = p_{R,21}^{*} = \rho_{R,1}\left(u_{R,1} - S_{R,1}\right)\left(u_{R,1} - u_I\right) + p_{R,1}$$

In case of a 1-2 contact, a similar approach is followed (Furfaro & Saurel, 2015).

The interfacial vairbales being locally constant on an element boundary during a time-step, system (19) can be rewritten under the following conservative form:

$$\frac{\partial \mathbf{Q}_k}{\partial t} + \frac{\partial \mathbf{F}_k(\mathbf{Q}_k)}{\partial x} = 0, \mathbf{Q}_k = \begin{pmatrix} \alpha_k \\ (\alpha\rho)_k \\ (\alpha\rho u)_k \\ (\alpha\rho E)_k \end{pmatrix}, \mathbf{F}_k(\mathbf{Q}_k) = \begin{pmatrix} \alpha_k u_I \\ (\alpha\rho u)_k \\ \alpha_k(\rho u^2 + P)_k - \alpha_k p_I \\ \alpha_k(\rho E + p)_k u_k - \alpha_k p_I u_I \end{pmatrix}$$

Local constancy of the interfacial variables also implies the decoupling of the full 7-wave Riemann problem in two 4-wave Riemann problems as shown in .

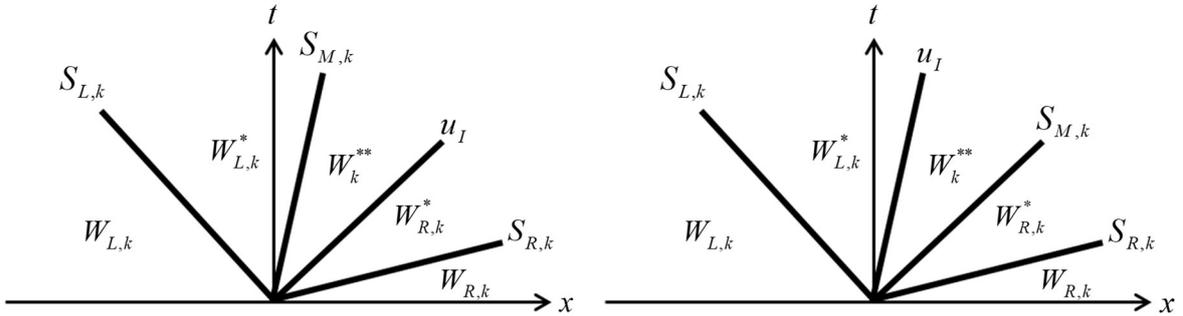

Figure 25: Main patterns of phase k Riemann problem: $S_{L,k} < S_{M,k} < u_I < S_{R,k}$ et $S_{L,k} < u_I < S_{M,k} < S_{R,k}$

The 4-wave Riemann solution is based on the Rankine-Hugoniot relations, approximating the wave speeds $S_{L,k}$ and $S_{R,k}$ by (Davis, 1988) and intermediate contact $S_{M,k}$ by HLL. Rankine-Hugoniot relations across the various waves read :

$$\mathbf{F}_{L,k}^{*} = \mathbf{F}_{L,k} + S_{L,k}\left(\mathbf{Q}_{L,k}^{*} - \mathbf{Q}_{L,k}\right)$$
$$\mathbf{F}_{R,k}^{*} = \mathbf{F}_{R,k} + S_{R,k}\left(\mathbf{Q}_{R,k}^{*} - \mathbf{Q}_{R,k}\right)$$
$$\mathbf{F}_{k}^{**} = \mathbf{F}_{R,k}^{*} + u_I\left(\mathbf{Q}_{k}^{**} - \mathbf{Q}_{R,k}^{*}\right)$$
$$\mathbf{F}_{L,k}^{*} = \mathbf{F}_{k}^{**} + S_{M,k}\left(\mathbf{Q}_{L,k}^{*} - \mathbf{Q}_{k}^{**}\right)$$

Explicit relations for $\mathbf{Q}^*_{L,k}, \mathbf{Q}^*_{R,k}$ and $\mathbf{Q}^{**}_k$ are obtained as a functions of known states $\mathbf{Q}_{L,k}$ and $\mathbf{Q}_{R,k}$. Riemann solution state is obtained through a classical sampling. Once ccomputed, these states and fluxes are used in the non-conservative explicit or implicit Godunov scheme associated to system (19).

The whole numerical flux function is algorithmically rather complex. Nevertheless, the ADOO procedure is able to provide the exact Jacobian matrix in a straightfoward manner.